\documentclass[manuscript]{aastex}

\usepackage[utf8]{inputenc}
\usepackage{longtable}

\shorttitle{Optical monitoring of BL Lac object S5 0716+714 and FSRQ 3C273 from 2000 to 2014}
\shortauthors{Yuan et al.}

\begin{document}

\title{Optical monitoring of BL Lac object S5 0716+714 and FSRQ 3C273 from 2000 to 2014}

\author{Yu-Hai Yuan$^{1,2}$, Jun-hui Fan$^{1,2}$, Jun Tao$^{3,4}$, Bo-Chen Qian$^{3,4}$, Denise Costantin$^{1,2}$, Hu-Bing Xiao$^{1,2}$, Zhi-Yuan Pei$^{1,2}$, Chao Lin$^{1,2}$}
\affil{$^{1}$Center for Astrophysics, Guangzhou University,
Guangzhou, Guangdong, 510006, PR.China.\\
$^{2}$Astronomy Science and Technology Research Laboratory of
   Department of Education of Guangdong Province, Guangzhou 510006,
   China.\\
$^{3}$Shanghai Astronomical Observatory, CAS, 80 Nandan Road,
Shanghai 200030, China\\
$^{4}$Joint Institute for Galaxies and Cosmology, ShAO and USTC,
CAS.
   }
\email{yh\_yuan@gzhu.edu.cn}

\begin{abstract}
% Context
{Using the 1.56m telescope at the Shanghai Observatory (ShAO), China, we monitored two sources, BL Lac object S5 0716+714 and Flat Spectrum Radio Quasar (FSRQ) 3C 273. For S5 0716+714, we report 4969 sets of CCD (Charge-coupled Device) photometrical optical observations (1369 for V band, 1861 for R band and 1739 for I band) in the monitoring time from Dec.4, 2000 to Apr.5, 2014. For 3C 273, we report 460 observations (138 for V band, 146 for R band and 176 for I band) in the monitoring time from Mar. 28, 2006 to Apr. 9, 2014.}
% Aims
{The observations provide us with a large amount of data to analyze the short-term and long-term optical variabilities. Based on the variable timescales, we can estimate the central black hole mass and the Doppler factor. An abundance of multi-band observations can help us to analyze the relations between the brightness and spectrum.}
% Method
{We use Gaussian fitting to analyze the intra-day light curves and obtain the intra-day variability (IDV) timescales. We use the discrete correlation
function (DCF) method and Jurkevich method to analyze the quasi-periodic variability. Based on the VRI observations, we use the linear fitting to analyze the relations between brightness and spectrum.}
% Results
{The two sources both show IDV properties for S5 0716+714. The timescales are in the range from 17.3 minutes to 4.82 hours; for 3C273, the timescale is $\Delta T = 35.6$ minutes. Based on the periodic analysis methods, we find the periods $P_V = 24.24\pm1.09$ days, $P_R=24.12\pm0.76$ days, $P_I=24.82\pm0.73$ days for S5 0716+714, and $P = 12.99 \pm 0.72$, $21.76 \pm 1.45$ yr for 3C273. The two sources displayed the "bluer-when-brighter" spectral evolution properties.}
% Conclusions
{S5 0716+714 and 3C 273 are frequently studied objects. The violent optical variability and IDV may come from the jet. Gaussian fitting can be used to analyze IDVs. The relations between brightness (flux density) and spectrum are strongly influenced by the frequency.}
\end{abstract}


\section{Introduction}
Blazars show some extreme properties, such as violently optical variability, core dominance, superluminal motion, and so on (Urry \& Padovani 1995; Ulrich, Maraschi \& Urry 1997). The optical variations of blazars last from minutes to years, and can be divided into two types: short-term and long-term variations. Being a special short-term variation, intra-day variabilities (IDV) show timescales from minutes to hours, and have been analyzed by many studies (Oke 1967; Miller 1975; Gupta et al.2008; Fan et al.2009a; Fan et al.2009b; Fan et al.2009c; Dai et al.2009; Poon et al.2009; Fan et al.2014; Yuan et al.2015a).

Blazars can be divided into two subclasses, BL Lacs and Flat Spectrum Radio Quasars (FSRQs),
the former being characterized by featureless optical
spectra or weak emission lines (Stickel et al.1991), and the latter
showing flat-spectrum radio spectra and typical
broad emission lines (Urry \& Padovani 1995). Generally, the division between the two subclasses is based on the equivalent
width (EW) of the optical broad emission; BL Lacs show $EW<5{\AA}$ (Urry \& Padovani 1995; Ghisellini et al.2011; Sbarrato et al.2012; Ghisellini \& Tavecchio 2015).

Time-scale is an important physical quantity, which is often used to probe the physical process of blazars. While the short-term variations, including intra-day variations, are usually non-periodic, the long-term ones are quasi-periodic.
In particular the first ones may come from the jets or the accretion disc. There are many theoretical models used to explain these variations, such as the shocks propagating from the relativistic jets (Marscher \& Gear 1985; Wagner \& Witzel 1995), hotspots or disturbances on or above accretion discs surrounding black holes (Chakrabarti \& Wiita 1993; Mangalam \& Wiita 1993).

%The short-term optical variations mightly come from the jets, accretion disc, and so on. There are many theoretical models used to explain this variations. For example, the shocks propagating from the relativistic jets (Marscher \& Gear 1985; Wagner \& Witzel 1995), hotspots or disturbances on or above accretion discs surrounding the black holes (Chakrabarti \& Wiita 1993; Mangalam \& Wiita 1993).

With an estimated redshift z=$0.31\pm0.08$ from the photometric detection of the
host galaxy (Nisson et al.2008), S5 0716+714 is a well known and frequently studied BL Lac object. It shows optical variabilities in the whole electromagnetic band with timescales from minutes to years. Intra-day variabilities (IDVs) have been studied by many authors (Poon et al.2009; Liu et al.2012; Gupta et al.2012; Bhatta et al.2013;  Man et al.2016; Lee et al. 2016, etc). Gupta et al.(2009) obtained timescales of 25 minutes; Rani et al.(2010) obtained timescales of 15 minutes; Man et al.(2016) obtained timescales of 17.6 minutes; and Bhatta et al.(2016) displayed the variability based on the peak-to-peak variations of 30\% and "bluer-when-brighter" spectral evolution.

%Bhatta et al.(2016)： Bhatta?G., Stawarz?L., Ostrowski?M., et al., 2016, ApJ, 831, 92

For this particular BL Lac object, many works have studied the time delays among different optical bands (Poon et al., 2009; Zhang 2010; Wu et al. 2012; Man et al. 2016). For example, Poon et al. (2009) found that the time delay between B and I band was $\tau_{BI}=11$ minutes; Zhang (2010) obtained time-delay values of a few minutes at different optical bands; Wu et al.(2012) got a time delay of 30 minutes between B and V bands and Man et al.(2016) obtained values of $1.308\pm0.603$ minutes between R and I bands and $1.445\pm0.511$ minutes between B and I bands.

Discovered
in 1963 by Smith \& Hoffleit (1963), 3C 273, with a redshift z=0.158, is one of the widely studied FSRQs and exhibits super-luminal motions (Unwin et al.1985). Asada et al.(2002) reported that 3C273 has helical magnetic structure. Dai et al.(2009) presented the long-term  $\emph{B, V, R, I}$ (BVRI)
observations and discussed the correlations between color index and brightness. Beaklini \& Abraham (2014) used the variability at 7mm band to find evidence of shocks and precession in the jet. Fan et al.(2014) analyzed the correlation between V-band flux density ($F_V$) and spectral index ($\alpha$), and found that $F_V=28$ mJy, suggesting two different correlations. When $F_V<28$ mJy, $F_V$ and $\alpha$ show anti-correlation, when $F_V>28$ mJy, $F_V$ and $\alpha$ show positive correlation. Yuan \& Fan (2015c) found an elliptic structure in the distribution of flux density and spectral indices, and that the time-span of the elliptic circle was consistent with the long-term optical periodicity of this source.

% Smith H.J. \& Hoffleit D., 1963, Nature, 198, 650
% Unwin S. C., Cohen M. H., Biretta J. A., et al., 1985, ApJ, 289, 109
% Asada K., Inone M. \& Uchida Y., 2002, PASJ, 54, 39
% Dai B.Z., Li X.H., Liu Z.M., et al., 2009, MNRAS, 392, 1181
% Beaklini P.P.B. \& Abraham Z., 2014, MNRAS, 437, 489
% Fan J. H., Kurtanidze O., Liu Y., et al.,2014, ApJS, 213,26
% Yuan Y.H. & Fan J.H., 2015, Ap\&SS, 357, 123

%As one of the mostly studied BL Lac object, S5 0716+714 show optical variabilities at the whole electromagnetic band with the timescales from minutes to years. The intra-day variabilities (IDVs) have been studied by many papers (Poon et al.2009; Liu et al.2012; Gupta et al.2012; Bhatta et al.2013;  Man et al.2016, etc). Based on optical observations, Gupta et al.(2009) obtained the timescales 25 minutes, then, Rani et al.(2010) obtained the timescales 15 minutes, and Man et al.(2016) obtained the timescales 17.6 minutes.

On the subject of blazars, there are many papers exploring the relationship between the spectrum and brightness (Edelson, Krolik \& Pike 1990; Trevese \& Vagnetti 2002; Vagnetti, Trevese \& Nesci 2003; Dai et al.2009; Poon, Fan \& Fu 2009; Yuan, Fan \& Pan 2015; Yuan \& Fan 2015). Stevens \& Gear (1999) analyzed the distributions of $\alpha_{ro}$ and $\alpha_{rx}$ (r, o, x, represents radio band, optical band and X-ray band, respectively), and obtained a correlation between the two spectral indices. Trevese et al.(2001) underlined a linear correlation between the variance of spectral index and the logarithmic flux density.

Generally, BL Lacs shows that the spectrum becomes flatter when the sources become brighter, and becomes steeper when the sources become fainter. However, FSRQs show a very complicated structure; some sources show a similar variation tendency to BL Lacs, and some sources show no variation tendency or flatter when the sources become fainter (Brown et al.1989; Carini \& Miller 1992; Fan et al.1998; Massaro, Nesci \& Maesano 1998; Nesci et al.1998; Speziali \& Natali 1998; Webb et al.1998; Fan 1999; Xie et al.1999; Xie et al.2002; Villata et al.2002; Gu et al.2006; Papadakis, Villata \& Raiteri 2007; Dai et al.2009; Poon, Fan \& Fu 2009; Yuan, Fan \& Pan 2015a; Yuan \& Fan 2015b).
In particular S5 0716+714 shows strong `bluer when brighter' correlations were not only found on timescales of one night but also during longer-term observations (Poon et al., 2009; Chandra et al. 2011; Wu et al. 2012; Man et al. 2016, etc).

This paper is arranged as follow: Section 2, presents our observations and data reductions; Section 3, presents our findings in terms of optical variability; Section 4, presents a discussion of our results in the context of current literature; and in Sections 5 we present our  conclusions.

\section{Observations and data reductions}

Observations were carried out using the Sheshan 1.56m optical telescope at the Shanghai Astronomical Observatory (ShAO), which is equipped with a cooled CCD camera (1024$\times$1024 pixels, 1 pixel=0.019 mm). The telescope view field is about 4'17" (1 pixel=0".25) and 13' (1 pixel=0".25) for a focal reducer. Another new liquid-nitrogen-cooled CCD camera (2048$\times$2048 pixels) has been used since August, 2002. The chip subtends to 11'$\times$11' in the sky and the scale is 0.31 arcsec per pixel (1 pixel=0.024 mm). The filters are standard Johnson B,V and Cousins R,I.

The bias images are taken at the beginning and the end of the night's observation. The flat-field images are taken at dusk and dawn.
The data reduction is carried out by the standard IRAF procedures based on the Linux system. Photometry is obtained after the bias, dark and flat-field correlations.

Given $K$, the number of comparison stars, for each of them ($S_{i}$, $i$=1,2,...K), we calculate the $i$th target magnitude ($m_i$): $m_i=m_{i|o}+m_{i|c}-m_{i|oc}$, here $m_{i|oc}$ is the observed comparison star magnitude, $m_{i|o}$ is the observed target magnitude, $m_{i|c}$ is the standard comparison star magnitude. Considering the whole comparison stars, the target magnitude ($m$) can be calculated as $m~=~\frac{\sum_{i=1}^{K}m_i}{K}$ with a standard error $\sigma~=~\sqrt{{\frac{\Sigma({m_i}-m)^{2}}{K-1}}}$.

In this work, for S5 0716+714, the VR comparison stars are from Villata et al.(1998) and the I comparison stars are from Ghisellini et al.(1997), and for 3C273, the comparison stars are from Smith et al.(1985). All the comparison stars have been listed in Table 1, in which, Col.1 is signed label of comparison stars, Col.2 is comparison stars at V band, Col.3 is comparison stars at R band, and Col.4 is comparison stars at I band.

\begin{table}
\caption{The comparison stars of S5 0716+714 and 3C 273}
\centering
\begin{tabular}{c|c|c|c|c}
\hline\hline
                  & Star   &    V(error)    &     R(error)       &      I(error)            \\
                  & (1)    &    (2)         &          (3)       &      (4)                 \\
\hline
S5 0716+714       &   A(2) & $11.46\pm0.01$ & $11.12\pm0.01$ & $10.92\pm0.04$  \\
                  &B(3)    & $12.43\pm0.02$ & $12.06\pm0.01$ & $11.79\pm0.05$  \\
                  &C(5)    & $13.55\pm0.02$ & $13.18\pm0.01$ & $12.85\pm0.05$  \\
                  &D(6)    & $13.63\pm0.02$ & $13.26\pm0.01$ & $12.97\pm0.04$  \\
\hline\hline
3C 273            & C      & $11.87\pm0.04$     & $11.30\pm0.04$ & $10.74\pm0.04$  \\
                  & D      & $12.68\pm0.04$     & $12.31\pm0.04$ & $11.99\pm0.06$  \\
                  & E      & $12.69\pm0.04$     & $12.27\pm0.05$ & $11.84\pm0.04$  \\
                  & G      & $13.56\pm0.05$     & $13.16\pm0.05$ & $12.83\pm0.05$  \\
\hline\hline
\end{tabular}
\end{table}

\section{Optical variability}\label{newmethod}

\begin{figure}
\begin{center}
\caption{The VRI light curves of S5 0716+714 (the upper sub-picture) and 3C 273 (the lower sub-picture). The black dots signify I band, the red dots signify R band and the green dots signify V band.}
\label{grad}
\end{center}
\end{figure}

For S5 0716+714 and 3C273, the V, R, I (VRI) light curves are shown in Fig.1, in which the upper panel displays S5 0716+714 and the lower panel 3C273.
For S5 0716+714, there are 4969 observations. At V band, there are 1369 observations, which cover $12.44\pm0.02$ mag to $15.11\pm0.01$ mag, with the averaged value $\overline{m_V}=13.51\pm0.42$ mag.
At R band, there are 1861 observations, which cover $11.67\pm0.04$ mag to $14.59\pm0.08$ mag, with the averaged value $\overline{m_R}=13.13\pm0.43$ mag.
At I band, there are 1739 observations, which cover $11.31\pm0.03$ mag to $14.09\pm0.02$ mag, with the averaged value $\overline{m_I}=12.57\pm0.41$ mag.

For 3C 273, there are 460 observations. At V band, there are 138 observations, which cover  $12.40\pm0.07$ mag to $13.38\pm0.07$ mag, with the averaged value $\overline{m_V}=12.73\pm0.15$ mag. At R band, there are 146 observations, which cover  $12.31\pm0.09$ mag to $12.88\pm0.12$ mag, with the averaged value $\overline{m_R}=12.57\pm0.09$ mag. At I band, there are 176 observations, which cover $11.78\pm0.06$ mag to $12.40\pm0.08$ mag, with the averaged value $\overline{m_I}=12.11\pm0.10$ mag.

\subsection{Intra-day optical variabilities}

In the intra-day light curves, if the optical variability $\Delta m > 3\sigma$, we use the Gaussian function to analyze the intra-day light curves, which can be explained by the following,
$$f(x)=A_0e^{\frac{-z^2}{2}}+A_3, z=\frac{x-A_1}{A_2},$$ where $A_0$, $A_1$, $A_2$, $A_3$ are the fitting parameter between $x$ and $f(x)$. In order to explain the four parameters, we take an example when $A_0=0.2$, $A_1=0.3$, $A_2=0.025$, and $A_3=13.5$. Fig.2 represents the change trend of $f(x)$ with $x$ ($A_0$, $A_1$, $A_2$, $A_3$ are known). If the change trend of $f(x)$ with $x$ can be fitted by a full Gaussian function, we identify this fitting profile as `w'; otherwise we identify it as `p'. If the intra-day light curves can be fitted by a full Gaussian function, the variable timescales $\Delta T$ can be calculated as: $\Delta T = 4 \times A_2$. Otherwise we take the actual span ($\Delta t$) of optical variability as the variable timescales $\Delta T$: $\Delta T$ = $\Delta t$.

\begin{figure}
\begin{center}
\caption{The change trend of $f(x)$ with $x$. $A_0$, $A_1$, $A_2$ and $A_3$ have been noted.}
\label{grad}
\end{center}
\end{figure}

For S5 0716+714, the IDV analyzed results are plotted in Fig.3 (I band), Fig.4 (R band), and Fig.5 (V band), in which, the red, blue, and green lines stand for the Gaussian fitting. Table 2 lists the main results, in which Col.1 is JD (+2450000);\
Col.2 is band;
Col.3 is $A_0$; Col.4 is $A_1$; Col.5 is $A_2$; Col.6 is $A_3$;
Col.7 is intra-day optical variability (in units of mag);
Col.8 is error, corresponding to Col.7 (in units of mag);
Col.9 is Gaussian function profile, `w': full Gaussian function, `p': part Gaussian function; and Col.10 is the IDV timescales.

Our results show that the IDV timescales are in the range 0.48 to 4.82 hours at I band, in the range 0.29 to 2.56 hours at R band, and in the range 0.38 to 2.56 hours at V band. The minimum timescales with the optical variability are $\Delta T_I$ =0.48 hours with $\Delta m=0.73\pm0.04$ mag on JD 2453388 at I band; $\Delta T_R$ =0.29 hours with $\Delta m=0.46\pm0.11$ mag on JD 2453774 at R band; and $\Delta T_V$ =0.38 hours with $\Delta m=0.35\pm0.10$ mag on JD 2453779 at V band.

\begin{figure*}
\begin{center}
%\includegraphics[width=0.194\textwidth]{I-IDV3050.eps}
%\includegraphics[width=0.194\textwidth]{I-IDV3051.eps}
%\includegraphics[width=0.194\textwidth]{I-IDV3388.eps}
%\includegraphics[width=0.194\textwidth]{I-IDV3727.eps}
%\includegraphics[width=0.194\textwidth]{I-IDV3774.eps}
%\\
%\includegraphics[width=0.194\textwidth]{I-IDV3779.eps}
%\includegraphics[width=0.194\textwidth]{I-IDV4122.eps}
%\includegraphics[width=0.194\textwidth]{I-IDV4131.eps}
%\includegraphics[width=0.194\textwidth]{I-IDV4166.eps}
%\includegraphics[width=0.194\textwidth]{I-IDV4809.eps}
%\\
%\includegraphics[width=0.194\textwidth]{I-IDV4818.eps}
\caption{For S5 0716+714, at I band, the results of intra-day light curves based on the Gaussian function. The red, green, and blue lines represent the Gaussian fitting curve.}
\label{trace}
\end{center}
\end{figure*}

\begin{figure*}
\begin{center}
\caption{For S5 0716+714, at R band, results of intra-day light curves based on the Gaussian function. The red, green, and blue lines represent the Gaussian fitting curve.}
\label{trace}
\end{center}
\end{figure*}

\begin{figure*}
\begin{center}
%\includegraphics[width=0.194\textwidth]{V-IDV3050.eps}
%\includegraphics[width=0.194\textwidth]{V-IDV3390.eps}
%\includegraphics[width=0.194\textwidth]{V-IDV3774.eps}
%\includegraphics[width=0.194\textwidth]{V-IDV3779.eps}
%\includegraphics[width=0.194\textwidth]{V-IDV4068.eps}
%\\
%\includegraphics[width=0.194\textwidth]{V-IDV4086.eps}
%\includegraphics[width=0.20\textwidth]{V-IDV4122.eps}
%\includegraphics[width=0.20\textwidth]{V-IDV4131.eps}
%\includegraphics[width=0.194\textwidth]{V-IDV4809.eps}
%\includegraphics[width=0.194\textwidth]{V-IDV4818.eps}
\caption{For S5 0716+714, at V band, results of intra-day light curves based on the Gaussian function. The red, green, and blue lines represent the Gaussian fitting curve.}
\label{trace}
\end{center}
\end{figure*}

\begin{table*}
\scriptsize
\caption{Result of IDV light curves analyzed by the Gaussian function of S5 0716+714}
\label{table}
\centering
\begin{tabular}{cccccccccc}
\hline
\hline
JD(2450000+)    &       Band    &       $A_0$   &       $A_1$   &       $A_2$   &        $A_3$    &       $\Delta m$      &       $\sigma$        & GF profile & $\Delta T$ \\
        &               &               &               &               &               &        /mag     &      /mag    &  & /hour \\
\hline
\hline
3050    &       I       &       0.699   &       3051.004        &       0.009   &       11.949  &        0.84     &       0.05    &      w       &        0.86   \\
3050    &       R       &       0.781   &       3051.009        &       0.012   &       12.366  &        0.786   &        0.114   &      w       &        1.15   \\
3050    &       V       &       0.539   &       3051.01 &       0.009   &       12.933  &        0.9      &       0.095   &       w       &       0.86    \\
\hline
3051    &       I       &       -0.277  &       3052.006        &       0.01    &       12.493  &        0.291   &        0.033   &      w       &        0.96   \\
3051    &       R       &       -0.323  &       3052.044        &       0.011   &       13.075  &        0.346   &        0.073   &      w       &        1.06   \\
\hline
3388    &       I       &       0.339   &       3388.136        &       0.029   &       12.339  &        0.706   &        0.04    &      w       &        2.78   \\
3388    &       I       &       -0.643  &       3388.207        &       0.005   &       12.537  &        0.734   &        0.041   &      w       &        0.48   \\
3388   &        R        &      0.863    &      3388.158        &       0.014   &       12.682  &        1.123    &      0.09     &      w    &  1.34  \\
\hline
3390    &       R       &       -0.281  &       3390.15 &       0.054   &       12.975  &        0.481    &       0.108   &       p      &        2.75   \\
3390    &       V       &       -1.244  &       3390.148        &       0.118   &       14.379  &        0.453   &        0.079   &      w       &        1.75   \\
3390    &       V       &       0.345   &       3390.222        &       0.019   &       13.074  &        0.453   &        0.079   &      w       &        1.82   \\
\hline
3702    &       R       &       0.278   &       3702.306        &       0.011   &       12.79   &        0.444   &        0.091   &      p       &        1.80   \\
\hline
3727    &       I       &       -0.138  &       3727.19 &       0.027   &       12.125  &        0.142    &       0.044   &       p      &        1.61   \\
\hline
3774    &       I       &       0.333   &       3774.119        &       0.015   &       12.828  &        0.573   &        0.095   &      w       &        1.44   \\
3774    &       R       &       0.35    &       3774.098        &       0.003   &       13.331  &        0.459   &        0.109   &      w       &        0.29   \\
3774    &       R       &       0.478   &       3774.193        &       0.012   &       13.372  &        0.554   &        0.135   &      p       &        2.16   \\
3774    &       V       &       0.388   &       3774.184        &       -0.06   &       13.639  &        0.868   &        0.053   &      p       &        2.54   \\
\hline
3779    &       I       &       -1.655  &       3778.836        &       0.145   &       13.336  &        0.262   &        0.057   &      p       &        1.48   \\
3779    &       R       &       0.27    &       3779.132        &       0.014   &       13.315  &        0.337   &        0.074   &      w       &        1.34   \\
3779    &   V   &    0.375      &   3779.062    &   0.027       &   13.647         &    0.300  &   0.126    &   w  &    2.59  \\
3779    &   V   &    0.381      &   3779.139    &   0.004       &   13.857         &    0.349      & 0.099          &   w  &    0.38  \\
\hline
4068    &   V   &   -0.256      &   4068.323  &         0.009   &   14.061      &    0.357        & 0.105 &        p  &    0.84   \\
\hline
4086    &       V       &       0.197   &       4086.261        &       0.015   &       13.979  &        0.198   &        0.054   &      w       &        1.44   \\
\hline
4122    &       I       &       -1.171  &       4122.286        &       0.095   &       12.253  &        0.306   &        0.047   &      p       &        1.70   \\
4122    &       V       &       0.293   &       4122.079        &       0.015   &       12.926  &        0.466   &        0.086   &      w       &        1.44   \\
\hline
4131    &       I       &       0.233   &       4131.148        &       0.032   &       12.35   &        0.366   &        0.087   &      w       &        3.07   \\
4131    &       V       &       0.202   &       4131.154        &       0.006   &       13.488  &        0.277   &        0.064   &      w       &        0.58   \\
\hline
4152    &       V       &       -0.234  &       4152.975        &       0.01    &       12.904  &        0.365   &        0.085   &      w       &        0.96   \\
\hline
4166    &       I       &       -2.552  &       4167.263        &       0.085   &       12.883  &        0.411   &        0.062   &      p       &        1.74   \\
\hline
4809    &       I       &       0.138   &       4809.419        &       0.072   &       12.198  &        0.114   &        0.038   &      p       &        4.82   \\
4809    &       V       &       0.167   &       4809.198        &       0.019   &       13.109  &        0.161   &        0.043   &      w       &        1.82   \\
4809    &       V       &       0.083   &       4809.288        &       0.024   &       13.126  &        0.07     &       0.016   &      w       &        2.30   \\
4809    &       V       &       -0.09   &       4809.34 &       0.022   &       13.225  &        0.074    &       0.023   &       w      &        2.11   \\
\hline
4818    &       I       &       -0.15   &       4818.143        &       0.025   &       12.417  &        0.155   &        0.058   &      w       &        2.41   \\
4818    &       I       &       -0.11   &       4818.235        &       0.019   &       12.476  &        0.126   &        0.032   &      w       &        1.82   \\
4818    &       R       &       0.602   &       4818.206        &       0.014   &       12.682  &        0.44     &       0.11    &      w       &        1.34   \\
\hline
4919    &       R       &       0.179   &       4920.01 &       0.007   &       13.882  &        0.327    &       0.068   &       w      &        0.67   \\
\hline
5170    &       R       &       0.219   &       5170.346        &       0.012   &       13.588  &        0.293   &        0.094   &      p       &        0.72   \\
\hline
\hline
\end{tabular}
\end{table*}

For 3C273, on Dec. 20, 2013 (JD 2456647), within 34.6 minutes, the optical variability is $\Delta m_I = 0.36\pm0.09$ mag, which is an obvious IDV; we refer to Fig.6, in which, the upper parts stand for the intra-day light curves and the lower parts stand for the magnitude difference between comparison stars `C' and `D'.

\begin{figure*}
\begin{center}
\caption{For 3C 273 at I band; results of intra-day light curves based on the Gaussian function. The red lines represent the Gaussian fitting curve.}
\label{trace}
\end{center}
\end{figure*}

\subsection{Quasi-periodic optical variability}

It's very important to choose suitable methods to analyze the long-term optical variability. Considering the uneven light curve, we use the discrete correlation
function (DCF) method and the Jurkevich method to deal with this question, and choose the common part as the quasi-periodicity.

The DCF method (Edelson \& Krolik 1988; Hufnagel \& Bregman 1992) can explore the correlation
from two variable temporal sets with a given time lag. If we only
input one set, we can calculate the period of the set. In order to
achieve this outcome, firstly we calculate the unbinned
correlation (UDCF) of the two data streams a and b, that is,
   \begin{equation}
      UDCF_{ij}=\frac{(a_{i}-\langle{a}\rangle)\times(b_{j}-\langle{b}\rangle)}{\sqrt{\sigma_{a}^{2}\times\sigma_{b}^{2}}},
   \end{equation}
where $a_{i}$, $b_{j}$ are two data streams, $\langle{a}\rangle$
and $\langle{b}\rangle$ are the average values of the data sets,
$\sigma_{a}$ and $\sigma_{b}$ are the corresponding standard
deviations. Secondly, we average the points through sharing the
same time lag by binning the $UDCF_{ij}$ in suitably sized
time-bins in order to get the DCF for each time lag $\tau$,
   \begin{equation}
     DCF(\tau)=\frac{1}{M}\sum UDCF_{ij}(\tau),
   \end{equation}
where M is the total number of pairs. The standard error of
each bin is
   \begin{equation}
     \sigma(\tau)=\frac{1}{M}(\sum[UDCF_{ij}-DCF(\tau)]^{2})^{0.5}.
   \end{equation}

The Jurkevich method (Jurkevich et al. 1971) is based on the expected mean
square deviation. It tests a run of trial periods around which the
data are folded. All data are assigned to m groups according to
their phases around each bin, and the whole $V_{m}^{2}$ for each
bin is computed; $V_{m}^{2}$ is the mean square deviation calculated by the square deviation of m groups. If the trial period is equal to the true one, then $V_{m}^{2}$ reaches its minimum. A `good' period will give a much
reduced variance relative to those given by `false' trial periods
and with almost constant values. The error can be determined by the
half width at half-maximum (HWHM) of the minimum in the $V_{m}^{2}$ , and $m=5$ is used in the following calculation.

For S5 0716+714, we use the DCF method to analyze the results, which are $P_{D|I}=24.82\pm0.73$ days, $P_{D|R} = 24.12\pm0.76$ days, $P_{D|V} = 24.24\pm1.09$ days ( Fig.7 (the upper three sub-pictures)). The results based on the Jurkevich method are $P_{J | I} = 27.28\pm2.68$ days, $P_{J | R} = 27.28\pm2.33$ days, $P_{J | V} = 26.62\pm2.01$ day (Fig.7 (the lower three sub-pictures)). Considering the common parts, we can obtain the quasi-periods, $P_V=24.24\pm1.09$ days (V band), $P_R=24.12\pm0.76$ days (R band), and $P_I=24.82\pm0.73$ days (I band).\\

\begin{figure*}
\begin{center}
%\includegraphics[width=0.33\textwidth]{I-DCF.eps}
%\includegraphics[width=0.33\textwidth]{R-DCF.eps}
%\includegraphics[width=0.33\textwidth]{V-DCF.eps}
%\\
%\includegraphics[width=0.33\textwidth]{I-Jurkevich.eps}
%\includegraphics[width=0.33\textwidth]{R-Jurkevich.eps}
%\includegraphics[width=0.33\textwidth]{V-Jurkevich.eps}
\caption{Using the Jurkevich and DCF methods to analyze the periodicities of VRI light curves. The left three panels show the DCF results and the right three panels show the Jurkevich results.}
\label{trace}
\end{center}
\end{figure*}

Fig. 8 shows our results from the use of the DCF
and Jurkevich methods to analyze the long-term variability for 3C 273. We cannot, however, obtain the periodic properties from these results. In order to analyze the long-term variability more comprehensively, we
collected the available data from the literature
(Angione \& Smith 1985, Fan et al.,2009, Dai et al.,2009,
Fan et al.,2014), supplemented our observations and then built the
light curves, which are shown in Fig.9.
Based on the DCF method, the results are as follows: $P_{D|1} =
13.20 \pm 1.63$ yr and $P_{D|2} = 22.69 \pm 2.28$ yr. Based on the Jurkevich
method, the results are $P_{J|1} = 0.99 \pm 0.13$ yr, $P_{J|2} = 2.05 \pm 0.23$ yr, $P_{J|3} = 12.99 \pm 0.72$ yr, and $P_{J|4} = 21.76 \pm 1.45$ yr. The calculated results are shown in
Fig.10. Considering the common parts, we can obtain the quasi-periods, $P_1 = 12.99 \pm 0.72$ yr and $P_2 = 21.76 \pm 1.45$ yr.

\begin{figure*}
\begin{center}
\caption{Calculation of the long-term periodicity of 3C273 based on our data. The upper panel shows the DCF result and the lower panel shows the Jurkevich result.}
\label{trace}
\end{center}
\end{figure*}

\begin{figure*}
\begin{center}
\caption{The light curves of 3C273. The black dots represent the literature data, and the red dots represent our observations.}
\label{trace}
\end{center}
\end{figure*}

\begin{figure*}
\begin{center}
\caption{The periodic results of 3C273. The first panel shows the DCF result with the red line being Gaussian fitting. The second panel shows the Jurkevich result.}
\label{trace}
\end{center}
\end{figure*}

\subsection{Optical spectral indices}

In order to obtain the optical spectral index ($\alpha$), firstly, we make the Galactic extinction correction; then, we convert the magnitude ($m_{\nu}$) into flux density ($F_{\nu}$); and lastly, we use the relation $F_{\nu}\propto\nu^{-\alpha}$ to calculate the spectral index ($\alpha$). Here $\nu$ is frequency.

1. We made the Galactic Extinction correction from NED (http://ned.ipac.caltech.edu/). For S5 0716+714, we use $A_V$ = 0.085$~mag$, $A_R$ = 0.067$~mag$, and $A_I$ = 0.047$~mag$; and for 3C 273, we use $A_V$ = 0.057$~mag$, $A_R$ = 0.045$~mag$, and $A_I$ = 0.031$~mag$.

2. To obtain the spectral index, we convert the magnitude ($m_{\nu}$) into flux density ($F_{\nu}$), here $\nu$ is frequency ($\nu$ = VRI, V: $6.17\times10^{14}$Hz, R: $4.78\times10^{14}$Hz, I:$3.89\times10^{14}$Hz).

3. VRI bands cannot expose at the same time; they in fact expose in turn. Considering the total exposure time of $\sim$1 minute for S5 0714+714, and $\sim$2 minutes for 3C273, we respectively choose 2 minutes for S5 0714+714 and 4 minutes for 3C273 as the maximum time span among three bands, and adopt the relation $F_{\nu}\propto\nu^{-\alpha}$ to calculate the spectral index ($\alpha$).

Converting the upper relation into the more convenient linear fitting form $log(F_{\nu})=-\alpha log(\nu)+const$, we input the VRI observations and calculate from this formula the $\alpha$ parameter. After calculation, there are 1095 spectral indices($\alpha$) in the range $0.16\pm0.09$ to $2.99\pm0.09$, with the averaged value $\overline{\alpha}=1.29\pm0.36$ for S5 0716+714. There are 85 spectral indices in the range  $0.17\pm0.09$ to $1.82\pm0.04$, with the averaged value $\overline{\alpha}=0.54\pm0.28$ for 3C 273.
\\

We use the linear fitting to analyze the relation between $\alpha$ and $F_{\nu}$ ($\nu$ = V, R, I): $F_{\nu}$ = $k\times\alpha$ + $b$. In this process, p is the chance probability of linear fitting, and r is the Pearson's correlation coefficient, which is expressed as (Press et al. 1994; Pavlidou et al. 2012; Fan et al. 2013): $$r=\frac{\Sigma(x_i-\overline{x})(y_i-\overline{y})}{\sqrt{\sum(x_i-\overline{x})^2}\sqrt{\sum(y_i-\overline{y})^2}},$$ where, $x_i$ is $F_{V,R,I}$, $y_i$ is $\alpha$, $\bar{x}$ is the averaged value of $\alpha$, and $\bar{y}$ is the averaged value of $F_{\nu}$ ($\nu$ = V, R, I).

For S5 0716+714, the correlations between $F_{\nu}$ ($\nu$ = V, R, I) and $\alpha$ are shown in Fig.11. At I band, $r_I=0.01$ and $p=83.8\%$; this result shows that there is no correlation. At R band, $\alpha=(-1.70\pm0.001)\times10^{-2}F_R+(1.62\pm0.006)$, with $r_R=-0.26$, $p = 4.51\times10^{-6}$; this result shows a strong anti-correlation. At V band,  $\alpha=(-3.51\pm0.001)\times10^{-2}F_V+(1.88\pm0.005)$, with $r_V=-0.48$, and $p = 2.56\times10^{-18}$; this result also shows a strong anti-correlation. With the frequency increasing ($I \rightarrow V$), the correlation shows the following variation: no correlation $\rightarrow$ strong anti-correlation.

For 3C 273, the relations between $F_{\nu}$ and $\alpha$ are shown in Fig.11 (the upper three panels.) At I band, $\alpha=(3.22\pm0.008)\times10^{-2}F_I-(0.65\pm0.11)$, with $r_I=0.37$, $p = 5.18\times10^{-4}$; this result shows a positive correlation. At R band, $r_R=-0.004$ and $p = 96.9\%$; this result shows no correlation. At V band, $\alpha=-(5.64\pm0.004)\times10^{-2}F_V+(2.24\pm0.04)$, with $r_V = -0.68$, $p = 4.94\times10^{-13}$; this result shows a strong anti-correlation.
With the frequency increasing ($I \rightarrow R \rightarrow V$), the correlations show the following variations: a positive correlation $\rightarrow$ no correlation $\rightarrow$ a strong anti-correlation.
\\

\begin{figure*}
\begin{center}
%\includegraphics[width=0.33\textwidth]{I-alf.eps}
%\includegraphics[width=0.33\textwidth]{R-alf.eps}
%\includegraphics[width=0.33\textwidth]{V-alf.eps}
%\\
%\includegraphics[width=0.33\textwidth]{alf-Fi.eps}
%\includegraphics[width=0.33\textwidth]{alf-Fr.eps}
%\includegraphics[width=0.33\textwidth]{alf-Fv.eps}
\caption{The correlations between $\alpha$ and $F_{IRV}$. The upper three panels show results for S5 0716+714, and the lower three panels shoiw results for 3C273. The red line represents the linear correlation between $\alpha$ and $F_{IRV}$.}
\label{trace}
\end{center}
\end{figure*}

\section{Discussion}

\subsection{IDV optical variability}

From the intra-day light curves of S5 0716+714 and 3C273, we can see that on individual days, the differences between the comparison stars are far from being random (Fig.3-6 (red dots)). The reason for this could be the strong atmosphere agitation during those days.

The IDV timescales of S5 0716+714 are in the range 17.3 minutes to 4.82 hours. We refer to Table 2, which can be divided into four parts, $T_1$, $T_2$, $T_3$ and $T_4$, as well as Fig.12. We use the Gaussian function to fit $T_1$, $T_2$, and $T_3$ and to obtain the averaged values $\overline{T_1} = 0.64 \pm0.05$ hours, $\overline{T_2} = 1.46 \pm0.04$ hours, and $\overline{T_3} = 2.52\pm0.02$ hours. $T_4$ has only one value, $T_4 = 4.82$ hours.

A large telescope on board the Fermi ($\textit{Fermi}$-LAT) supplies us with abundant $\gamma$-ray data of blazars, among which there are many sources that show lares with variability timescales of about $10^4$ s (Abdo et al.2009, Abdo et al.2010a, Abdo et al.2010b, Ackermann et al. 2010, Tavecchio et al.2010, Foschini et al.2011). Our timescales are in the range of $0.64\sim4.82$ hours ($2.3\times10^3\sim1.7\times10^5$ s). These are consistent with the $\gamma$-ray flare timescales, which suggests that the optical IDVs and $\gamma$-ray flares might come from the same place in the jet of S5 0716+714.

% Fossati G., Maraschi L., Celotti A., Comastri A., Ghisellini G., 1998, MNRAS, 299, 433

\begin{figure*}
\begin{center}
\caption{The $\Delta T$ distributions of S5 0716+714, which can be divided into three parts, fitted by Gaussian function (red lines).}
\label{trace}
\end{center}
\end{figure*}

\subsection{The central-black-hole mass}

In blazars, the central black holes play a very important role in the observational properties and attract great attention. The black hole mass might shed some light on the evolution of active galactic nuclei (AGNs) (Fan 2005) and there are many methods to calculate this parameter, such as velocity dispersion (Wu et al. 2002, Woo \& Urry 2002, Woo et al., 2005, Sbarrato et al. 2012), reverberation mapping (Woo \& Urry 2002), optical luminosity (Kawakatu et al. 2007, Zhou \& Cao 2009),  and so on.

The origin of short timescales are probably produced in the innermost part of the blazar and the region near the black hole, such as the accretion disc, the broad line region, etc. In this sense, the short-term timescale can be used to estimate the mass (Abramowicz \& Nobili 1982; Miller et al. 1989).
For S5 0716+714, we use the
Gaussian function to analyze the intra-day light curves, and obtain
a minimum timescale of $\Delta T = 17.3$ minutes. For 3C
273, the IDV timescale obtained is $\Delta T = 35.6$ minutes.

If in the center of the thin accretion disks lies a Schwarzschild black hole, the radius is $r_s = \frac{6GM}{c^2}$. But if there lies
a Kerr black hole, the radius is $r_k = 1.48\times10^5(1+\sqrt{1-a^2})\frac{M}{M_{\odot}}$ ,
where a is the angular momentum parameter (Witta 1985). Consider
the period $p = \frac{2\pi{r}}{c}$, that is, $r = 4.8\times10^9\frac{p}{1+z}$, $p$ in units of second. For a Schwarzschild black hole, the mass can be $M_s = 5.3\times(\frac{p}{1+z})\times10^3M_{\odot}$ ($M_{\odot}$: the Solar mass). For a Kerr black hole (a=1), the mass can be $M_k = 3.2\times10^4\times(\frac{p}{1+z})M_{\odot}$. `We use the short-term timescale $\Delta t$ to indicate the period `p'.

For S5 0716+714, the redshift z=$0.31\pm0.08$ (Nisson et
al.2008) and $\Delta T = 17.3$ minutes. Furthermore, we obtained the central black hole:
$M_s = 4.2 \times 10^6M_{\odot}$ for a Schwarzschild black hole and $M_k =
2.56\times10^7M_{\odot}$ for a Kerr black hole. Gupta et al.(2009) used IDV timescale to obtain the mass $2.5\times10^6M_{\odot}$, which is very consistent with our results. Liang \& Liu (2003) used the optical luminosity to obtain $M = 1.25 \times 10^8M_{\odot}$, which is heavier than our results. The reason for this might be that our value gives the lower limit.

For 3C 273, z = 0.158, and therefore the black hole mass is $9.5\times10^6M_{\odot}$
for a Schwarzschild black hole and $5.74\times10^7M_{\odot}$ for a Kerr black
hole. Espaillat et al.(2008) obtained a black hole mass of about
$8.1\times10^7M_{\odot}$, which is consistent with our result. Paltani \& Turler (2005) applied the reverberation method
to find a maximum-likelihood mass of about $6.59\times10^9M_{\odot}$, which is heavier than ours and Espaillat
et al.(2008)'s results.

% Wu X.B., Liu, F.K., Zhang, T.Z., 2002, A\&A, 389, 742
% Woo J.H. \& Urry C.M., 2002, AJ, 579, 530
% Fan J.H., 2005, A\&A, 436, 799
% Woo J.H., Urry C.M.,van der Marel Roeland P., et al., 2005, ApJ, 631, 762
% Kawakatu, N, Imanishi M, Nagao T, 2007, ApJ, 661, 660
% Zhou M, Cao X.W,2009, RAA, 9, 293Z
% Sbarrato, T., et al., 2012, MNRAS, 421, 1764

% Becker P., Kafatos M., 1995, ApJ 453, 83
% Bednarek W., 1993, A\&A 278, 307
% Cheng K.S., Fan J.H., Zhang L. A\&A, 1999, 352, 32-48
% Fan J.H., 2005, A\&A, 436, 799
% Abdo, A.A., Ackermann, M., Atwood, W.B., et al., 2009, ApJ, 697, 934
% Abdo, A.A., Ackermann, M., Ajello, M., et al., 2010a, ApJ, 710, 810
% Abdo, A.A., Ackermann, M., Ajello, M., et al., 2010b, ApJ, 714, L73
% Ackermann, M., Ajello, M., Baldini, L., et al., 2010, ApJ, 721, 1383
% Tavecchio, F., Ghisellini, G., Bonnoli, G., Ghirlanda, G. 2010, MNRAS, 405, L94
% Foschini et al. 2011, A\&A, 530, 77
%

%$\lambda=0.1$, $M=9.95 M_7$, $\Phi=42.29^{\circ}$, $d=41.30 R_g$ ($R_g$: the Schwarzschild radius), $\delta=1.56$
%$\lambda=1.0$, we can calculate $M=6.99 M_7$, $\Phi=46.73^{\circ}$, $d=43.30 R_g$, $\delta=1.15$

% Acero,?F., Ackermann,?M., Ajello,?M., et al., 2015, ApJS, 218, 23

% when $\lambda=0.1$, $\Phi = 53.7^{\circ}$, $d = 33.67 R_g$, $\delta = 0.74$
% when $\lambda=1.0$, $\Phi=57.11^{\circ}$, $d=35.25 R_g$, $\delta = 0.56$

\begin{figure*}
\begin{center}
\caption{The spectral variation of 3C 273 and S5 0716+714. The denser distributions are indicated by the rectangle regions. }
\label{trace}
\end{center}
\end{figure*}

\begin{figure*}
\begin{center}
%\includegraphics[width=0.325\textwidth]{aFI2910.eps}
%\includegraphics[width=0.335\textwidth]{aFR2910.eps}
%\includegraphics[width=0.33\textwidth]{aFV2910.eps}
%\\
%\includegraphics[width=0.33\textwidth]{aFI3305.eps}
%\includegraphics[width=0.33\textwidth]{aFR3305.eps}
%\includegraphics[width=0.33\textwidth]{aFV3305.eps}
%\\
%\includegraphics[width=0.325\textwidth]{aFI3600.eps}
%\includegraphics[width=0.335\textwidth]{aFR3600-3.eps}
%\includegraphics[width=0.325\textwidth]{aFV3600-3.eps}
%\\
%\includegraphics[width=0.325\textwidth]{aFI4000.eps}
%\includegraphics[width=0.33\textwidth]{aFR4000.eps}
%\includegraphics[width=0.33\textwidth]{aFV4000.eps}
%\\
%\includegraphics[width=0.33\textwidth]{aFI4800.eps}
%\includegraphics[width=0.33\textwidth]{aFR4800.eps}%
%\includegraphics[width=0.33\textwidth]{aFV4800.eps}
\caption{For S5 0716+714, the relations between $\alpha$ and $F_{VRI}$ at the five denser regions. The black dots and the red dots represent the different parts of the same regions. The green lines stand for the linear fitting between $\alpha$ and $F_{VRI}$.}
\label{trace}
\end{center}
\end{figure*}

\begin{figure*}
\begin{center}
%\includegraphics[width=0.325\textwidth]{aFI-3C273-1.eps}
%\includegraphics[width=0.335\textwidth]{aFR-3C273-1.eps}
%\includegraphics[width=0.325\textwidth]{aFV-3C273-1.eps}
%\\
%\includegraphics[width=0.33\textwidth]{aFI-3C273-2.eps}
%\includegraphics[width=0.33\textwidth]{aFR-3C273-2.eps}
%\includegraphics[width=0.33\textwidth]{aFV-3C273-2.eps}
%\\
%\includegraphics[width=0.33\textwidth]{aFI-3C273-3.eps}
%\includegraphics[width=0.33\textwidth]{aFR-3C273-3.eps}
%\includegraphics[width=0.33\textwidth]{aFV-3C273-3.eps}
\caption{For 3C273, the relations between $\alpha$ and $F_{VRI}$ at the three denser regions. The black dots and the red dots stand for the different parts of the same regions. The green lines show the linear fitting between $\alpha$ and $F_{VRI}$.}
\label{trace}
\end{center}
\end{figure*}

\subsection{Doppler factor}

The Doppler factors of S5 0716+714 and 3C 273 have been studied by many authors (Ghisellini et al. 1993; Hartman et al.1999; Fan et al.2009a; Hovatta et al.2009; Savolainen et al.2010; Fan et al.2014).

Based on the relativistic beaming model, the optical depth can be calculated from the pair-production. Based on the work of Mattox et al.(1993), Fan et al. (2013) deduced the lower limit of the Doppler factor ($\delta$),
\begin{equation}
\delta\geq[1.54\times10^{-3}(1+z)^{4+2\alpha_X}(\frac{d_L}{Mpc})^2(\frac{\Delta T}{h})^{-1}(\frac{F_{keV}}{\mu{Jy}})(\frac{E_{\gamma}}{GeV})^{\alpha_X}]^{\frac{1}{4+2\alpha_X}},
\end{equation}
where $F_{keV}$ is X-ray flux at 1keV (in unit of $\mu$Jy), $\alpha_X$ is X-ray spectral index, $d_L$ is luminosity distance (in unit of Mpc), $\Delta{T}$ is timescale (in unit of hour),  and $E_{\gamma}$ is averaged $\gamma$-ray photon energy (in unit of GeV).

For S5 0716+714, $d_L = 1499$ Mpc (NED, http://ned.ipac.caltech.edu/), $F_{keV}=0.99\mu$Jy , $\alpha_X = 1.77$ (Donato et al. 2001), and $E_{\gamma}~=~4.59$ GeV (Fan et al. 2014b). Our minimum $\Delta{T}$ is $0.64$ hours, so we can obtain $\delta \geq 5.89$.
There are many works dealing with the Doppler factor of S5 0716+714, for example, Ghisellini et al.(1993): $\delta=2.1$, Hovatta et al.(2009): $\delta = 10.9$, Fan et al.(2009d): $\delta = 8.76$ and Savolainen et al.(2010): $\delta = 10.8$. Our results are consistent with those by Fan et al.(2009d), Hovatta et al (2009), and Savolainen et al (2010).

For 3C 273, $d_L~=~734$ Mpc (NED), $F_{keV}=10.921\mu$ Jy,  $\alpha_X = 2.11$ (Brinkmann et al.1997), $E_{\gamma}~=~2.82$GeV (Fan et al. 2014b), and $\Delta{T}=34.6/60$ hour, so we can obtain $\delta\geq6.64$. Hartman et al.(1999) obtained $\delta=6$. Fan et al.(2009c) used the radio optical timescales and obtain $\delta=6.05$. Zhang et al.(2013) obtained $\delta = 7.4\pm0.9$; Hovatta et al.(2009) obtained $\delta = 17$ ; and Savolainen et al.(2010) obtained $\delta =  16.8$. Our results are consistent with those works.

\begin{table*}
\scriptsize
\caption{The linear correlations between $\alpha$ and $F_{IRV}$ of S5 0716+714 and 3C273}
\label{table}
\centering
\begin{tabular}{c|c|c|c|c|c|c|c}
\hline
\hline
Name    & Span &        $\alpha-F_{\nu}$        &       $k\times10^{-2}$        &       $b$      &        $r$    &       $p$      &      note    \\
(1)     & (2) & (3)     &       (4)     &       (5)     &       (6)     &       (7)      &       (8)     \\
\hline
\hline
S5 0716+714     &       gross   &       $\alpha-F_{I}$  &       -       &       -       &       0.01     &        83.8\% &        No Corr        \\
        &               &       $\alpha-F_{R}$  &       $-1.70\pm0.001$ &       $1.62\pm0.006$   &        -0.26  &        $4.51\times10^{-6}$    &       Strong Anti-Corr        \\
        &               &       $\alpha-F_{V}$  &       $-3.51\pm0.001$ &       $1.88\pm0.005$   &        -0.48  &        $2.56\times10^{-18}$   &       Strong Anti-Corr        \\
\hline
        &       JD2910-JD3079   &       $\alpha-F_{I}$  &       -       &       -       &       0.12     &        30.3\% &        No Corr        \\
        &       (region-1)      &       $\alpha-F_{R}$  &       $-1.66 \pm 0.006$  &        $1.63 \pm 0.04$        &        -0.25  &       $3.59\times10^{-2}$     &       Anti-Corr       \\
        &               &       $\alpha-F_{V}$  &       $-4.34 \pm0.005$        &       $2.11 \pm 0.03$        &       -0.56  &        $6.16\times10^{-7}$    &       Strong Anti-Corr        \\
\hline
        &       JD3305-JD3443   &       $\alpha-F_{I}$  &       $-2.10 \pm 0.01$   &        $2.08 \pm 0.09$        &        -0.28  &       $5.85\times10^{-2}$     &       Weak Anti-Corr        \\
        &       (region-2)      &       $\alpha-F_{R}$  &       $-5.21 \pm 0.03$   &        $2.56 \pm 0.05$        &        -0.58  &       $3.21\times10^{-5}$     &       Strong Anti-Corr        \\
        &               &       $\alpha-F_{V}$  &       $-6.94 \pm 0.01$        &       $2.62 \pm 0.03$        &       -0.69  &        $1.50\times10^{-10}$   &       Strong Anti-Corr        \\
\hline
        &       JD3670-JD3811   &       $\alpha-F_{I}$  &       $-1.07 \pm 0.006$  &        $1.53 \pm 0.006$        &      -0.24   &       $2.54\times10^{-5}$     &       Anti-Corr \\
        &       (region-3)      &       $\alpha-F_{R}$  &       $-2.09 \pm 0.002$  &        $1.71 \pm 0.004$       &        -0.42  &       $1.75\times10^{-14}$    &       Strong Anti-Corr        \\
        &               &       $\alpha-F_{V}$  &       $-3.29 \pm 0.002$       &       $1.86 \pm 0.003$       &      -0.61    &      $1.10\times10^{-31}$    &       Strong Anti-Corr        \\
\hline
        &       JD4068-JD4173   &       $\alpha-F_{I}$  &       $-3.84 \pm 0.004$  &        $1.41 \pm 0.004$        &      -0.11   &       $7.8\times10^{-2}$      &       Weak Anti-Corr        \\
        &       (region-4)      &       $\alpha-F_{R}$  &       $-1.29 \pm 0.006$  &        $1.57 \pm 0.002$       &        -0.31  &       $1.63\times10^{-7}$     &       Strong Anti-Corr       \\
        &               &       $\alpha-F_{V}$  &       $-2.19 \pm 0.0008$      &       $1.67 \pm 0.002$       &      -0.43    &      $2.99\times10^{-13}$    &       Strong Anti-Corr        \\
\hline
        &       JD4796-JD5284   &       $\alpha-F_{I}$  &       $-5.38 \pm 0.0007$ &        $1.56 \pm 0.004$        &      -0.15   &       $4.15\times10^{-2}$     &       Weak Anti-Corr        \\
        &       (region-5)      &       $\alpha-F_{R}$  &       $-1.31\pm0.001$ &       $1.66 \pm 0.003$      &        -0.28  &       $8.55\times10^{-5}$     &       Anti-Corr       \\
        &               &       $\alpha-F_{V}$  &       $-2.14 \pm 0.001$       &       $1.73 \pm 0.003$       &      -0.38    &      $7.62\times10^{-8}$     &       Strong Anti-Corr       \\
\hline
\hline
3C 273  &       gross   &       $\alpha-F_{I}$  &       $3.22 \pm 0.008$        &        $-0.65 \pm 0.11$        &        0.37   &       $5.18\times10^{-4}$     &       Strong Corr     \\
        &               &       $\alpha-F_{R}$  &       -       &       -       &       -0.004  &       96.9\%   &        No Corr        \\
        &               &       $\alpha-F_{V}$  &       $-5.64 \pm 0.004$       &       $2.24 \pm 0.04$        &       -0.68  &        $4.94\times10^{-13}$   &       Strong Anti-Corr        \\
\hline
        &       JD4500-JD4900   &       $\alpha-F_{I}$  &       $5.32\pm0.02$   &        $-1.39\pm0.23$  &        0.68    &      $5.08\times10^{-4}$     &       Strong Corr    \\
        &       (region-1)      &       $\alpha-F_{R}$  &       $6.76\pm0.03$   &        $-1.44\pm0.28$  &       0.65     &      $1.02\times10^{-3}$     &       Strong Corr    \\
        &               &       $\alpha-F_{V}$  &       $-6.15\pm0.04$  &       $2.43\pm0.36$    &        -0.56  &        $6.03\pm10^{-3}$       &       Strong Anti-Corr        \\
\hline
        &       JD5245-JD5500   &       $\alpha-F_{I}$  &       -       &       -       &       0.15     &        43\%   &        No Corr        \\
        &       (region-2)      &       $\alpha-F_{R}$  &       $-7.97\pm0.04$  &        $2.86\pm0.36$    &      -0.59    &      $5.95\times10^{-4}$     &       Strong Anti-Corr        \\
        &               &       $\alpha-F_{V}$  &       $-6.39\pm0.008$ &       $2.37\pm0.07$    &        -0.80  &        $1.13\times10^{-7}$    &       Strong Anti-Corr        \\
\hline
        &       JD5500-JD5700   &       $\alpha-F_{I}$  &       $4.21\pm0.05$   &        $-1.03\pm0.73$  &        0.37    &      $8.53\times10^{-2}$     &       Weak Corr    \\
        &       (region-3)      &       $\alpha-F_{R}$  &       $-6.58\pm0.09$  &        $2.51\pm0.87$    &      -0.43    &      $4.43\times10^{-2}$     &       Anti-Corr       \\
        &               &       $\alpha-F_{V}$  &       $-6.61\pm0.01$  &       $2.52\pm0.13$    &        -0.78  &        $1.77\times10^{-5}$    &       Strong Anti-Corr        \\
\hline
\hline
\end{tabular}
\end{table*}

\subsection{Relations between brightness and spectrum}

\subsubsection{S5 0716+714}

For S5 0716+714, at different bands, $\alpha$ and $F_{VRI}$ show different relations. At V and R band, there exists strong anti-correlation; at I band, there exists no-correlation. We check the spectral variation, and find that there are five denser regions, which have been noted by the rectangular boxes 1, 2, 3, 4 and 5 ( Fig.13 (the left panel)).

For the five regions, we analyzed their correlations (Fig.14). The analyzed results have been listed in Table 3 (Col.5 displays the intercept, b, Col.6 the correlation coefficient, and Col.7: the chance probability. At I band, on most of the regions, $\alpha$ and $F_{I}$ show no correlation or weak anti-correlation (except for region-3), and $\alpha$ and $F_{I}$ show anti-correlation. At R band, on Regions 1 and 5 there is anti-correlation, and on Regions 2, 3, and 4, $\alpha$ and $F_{R}$ show strong anti-correlation. At V band, on the whole five Regions, $\alpha$ and $F_{V}$ show strong anti-correlations. Comparison with the correlations at three bands shows that with the frequency increasing ($I\rightarrow{R}\rightarrow{V}$), the correlations tend to be strong anti-correlation, which is consistent with the variation tendency from the gross sample.

In some regions, with the flux densities increasing, $\alpha$ and $F_{VRI}$ show different relations. For example, in Region-2, at I band, there is a break point, $F_{I|b} = 27 mJy$. When $F_{I} < 27 mJy$, no correlation exists between $\alpha$ and $F_{I}$; but when $F_{I} > 27 mJy$, there is strong correlation, $\alpha=(0.09\pm0.01)F_I-(1.68\pm0.66)$, with r=0.68, $p=1.0\times10^{-4}$. In Region-3, the break points are $F_{I|b} = 29 mJy$, $F_{R|b} = 23 mJy$ and $F_{V|b} = 14.5 mJy$. In Region-4, the break points at three bands are $F_{I|b} = 31.5 mJy$ , $F_{R|b} = 19.3 mJy$, $F_{V|b} = 20 mJy$. In Region-5, the break points at three bands are $F_{I|b} = 24 mJy$, $F_{R|b} = 16 mJy$, and $F_{V|b} = 12 mJy$. The break points at each region have been noted in Fig.14. The upper results show that the relations between spectrum and brightness can be influenced by the brightness of the source.

\subsubsection{3C 273}

For 3C273, at different bands, $\alpha$ and $F_{VRI}$ show different relations. From I to V band, the relations are from strong correlation to no correlation to strong anti-correlation. The spectral variation of this source has been shown in Fig.13 (the right panel), based on which, we can find three denser regions, noted by the rectangular boxes 1, 2, and 3.

The analyzed results concerning the three regions are listed in Table 3 and shown in Fig.15. At I and R band, there exist different relations, from strong correlation to strong anti-correlation. At V band, in each of these three Regions, $\alpha$ and $F_{V}$ show strong anti-correlations. Comparing these results with the whole sample, we find that with increasing frequency ($I\rightarrow{R}\rightarrow{V}$), the correlations tend to be strong anti-correlation, which is consistent with the variation tendency of the sample as a whole.

S5 0716+714 and 3C 273 belong to different subclasses of Blazars. At V band, both objects show "bluer-when-brighter" evolution properties. At R and I band, the relations between brightness and spectrum are not fixed.
Fiorucci, Ciprini \& Tosti (2004) pointed out that the optical spectrum of QSOs consists of two components; one variable ($\alpha_{v}$, with a flatter slope), which comes from synchrotron emission, and the other part stable ($\alpha_{s}$), which might come from the thermal emission.
For S5 0716+714 and 3C273, at V band, most emissions could possibly come from synchrotron emission, and the relations between $F_{V}$ and $\alpha$ show anti-correlation. With decreasing detection frequency, the near IR emission becomes important, and $\alpha-F_{RI}$ show complicated results.

\subsection{Quasi-periods}

There are many sources showing quasi-periods, such as 3C273, 3C279, 3C454.3, Mrk335, Mrk 421, Mrk 501, OJ287, 0109+224, 0735+178, 2200+420, and so on (Kunkel 1967; Sillanpaa et al.1988; Chertoprud et al.1973; Liu et al.1995; Stickel et al.1993; Fan et al.1998; Fan \& Lin 2000; Xie et al.2002; Raiteri et al.2001; Ciprini et al.2003; Fan et al. 2007, 2014). Explanation of the long-term variations could be based on the binary black-hole model, the thermal instability model or the perturbation model, and so on (Fan et al.2007).

The quasi-periods of 3C 273 have been studied frequently. Smith \& Hoffleit (1963) analyzed the optical light curve of 1887-1963 and obtained a period of $12.7\sim15.2$ yr. Babadzhanyants \& Belokon (1993) found a period of 13.4 yr. Based on 110 yr of optical data, Fan et al.(2001) obtained periods of $\sim$2.0, $13.65\pm0.20$, and $22.5\pm0.2$ yr. Vol'vach et al.(2013) claimed periods of $2.8 \pm 0.3$, $4.9 \pm 0.3$ yr, $7.2 \pm 0.8$, $11.2 \pm 2.3$ yr in optical band. Our result $P_1 = 12.99 \pm 0.72$ yr is consistent with Smith \& Hoffleit (1963), Babadzhanyants \& Belokon (1993), Fan et al.(2001), Vol'vach et al.(2013), and the other result of $P_2 = 21.76\pm1.46$ yr  was also found by Fan et al.(2001).

When we use the Jurkevich method to analyze the long-term variation, we find the other two results $P_{J} = 0.99 \pm 0.13$, $2.05 \pm 0.23$. These results, however, are not shown when using DCF methods, and therefore should not be taken as quasi-periods. The Jurkevich results might be influenced by the sampling because 3C 273 is close to the ecliptic and cannot be observed all year round as the Sun is sometimes too close.

%Smith H. J., \& Hoffleit D. 1963, Natur, 198, 650
%Babadzhanyants M. K., \& Belokon E. T. 1993, ARep, 37, 127

\section{Conclusion}\label{dyn}

In this work, we present VRI photometric results of S5 0716+714 and 3C273, which were obtained using the 1.56m telescope at the Shanghai Observatory. Based on these observations, we come to the following conclusions.

Although S5 0716+714 and 3C 273 belong to different sub-classes of Blazars, both sources show complex dependency on the spectral index and flux density. The relations between these two parameters can be strongly influenced by the frequency and brightness.

Furthermore, the timescales provided here could be used to constrain some important physical parameters, such as the black hole, Doppler factor, emitting region, and so on.
The causes of IDV and long-term optical periodicity are not clear and definite. To answer this question, further observations and improvement of the emitting theory are required. Gaussian function fitting is very useful when we calculate the timescales and study the whole variation trend of optical variability.

\acknowledgments
We thank the anonymous referee for useful comments. The work is partially supported by the National Natural Science
Foundation of  China (NSFC 11403006, USFC U1531245, NSFC 10633010, NSFC 11173009 and NSFC U1431112), Science and Technology Program of Guangzhou (201707010401), Guangdong Province Universities and Colleges Pearl River Scholar Funded
Scheme(GDUPS 2009), Yangcheng Scholar Funded Scheme(10A027S), Innovation team in Guangdong Province (2014KCXD014) and
the finaincial support for the Key subject of Guangzhou City.

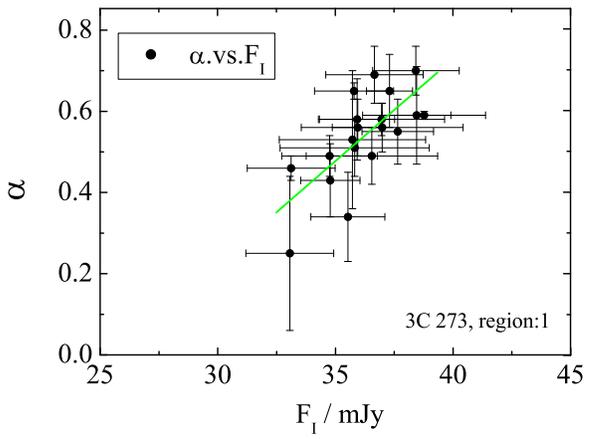
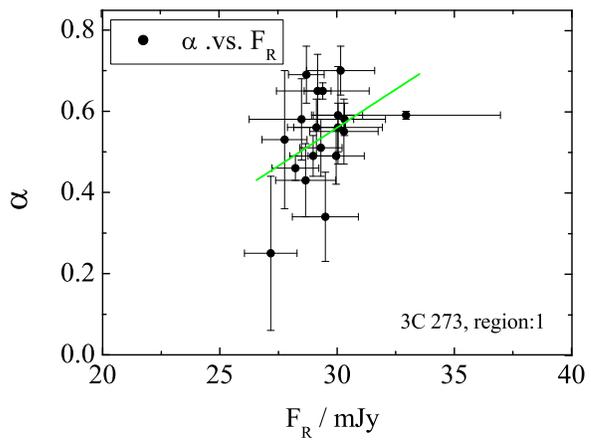
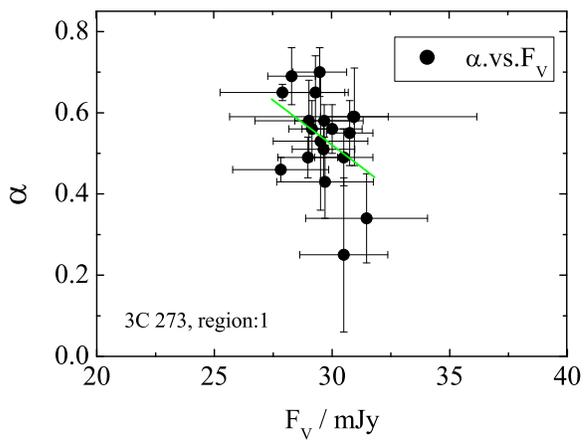
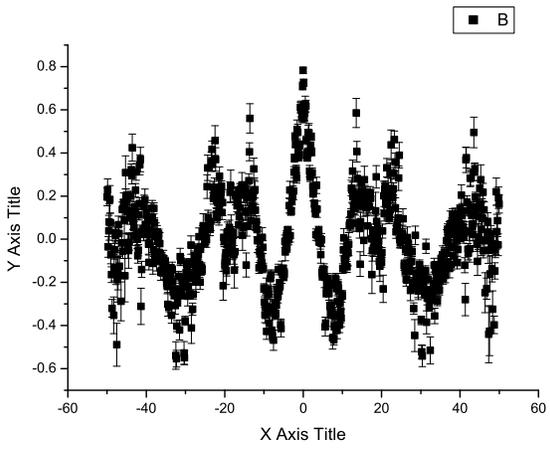
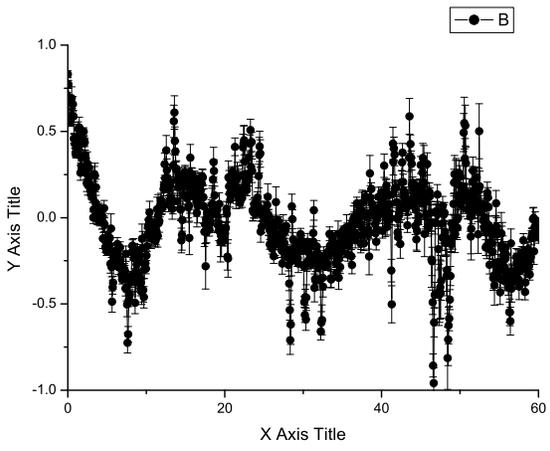
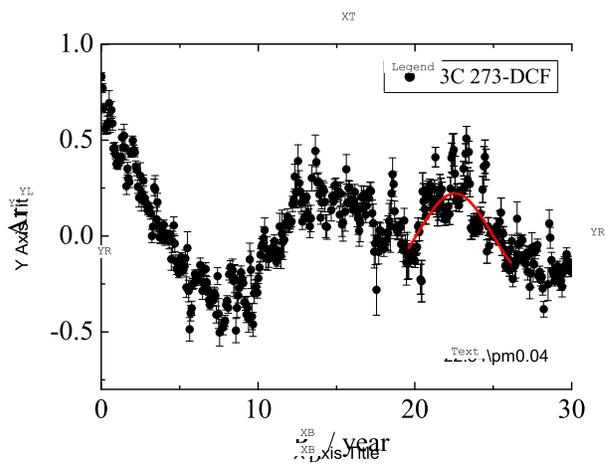
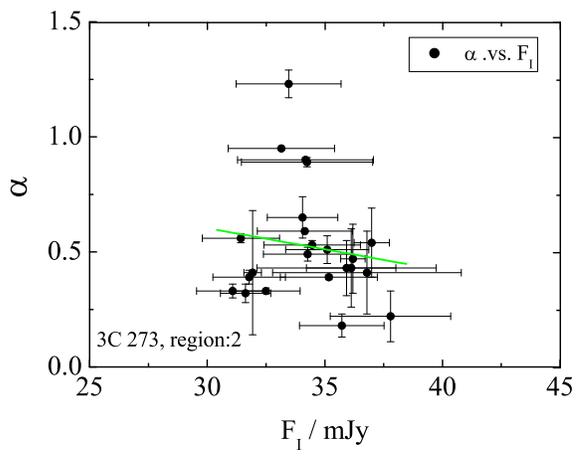
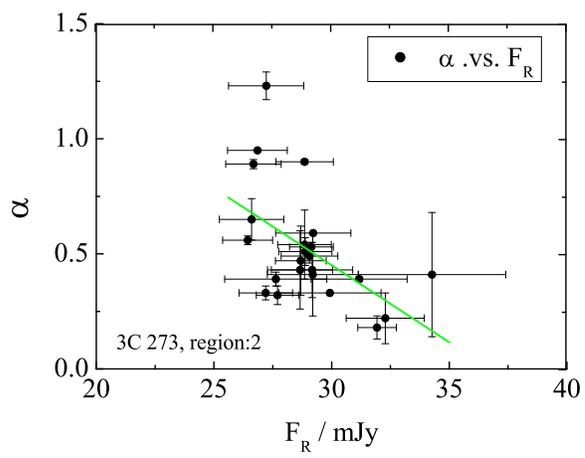
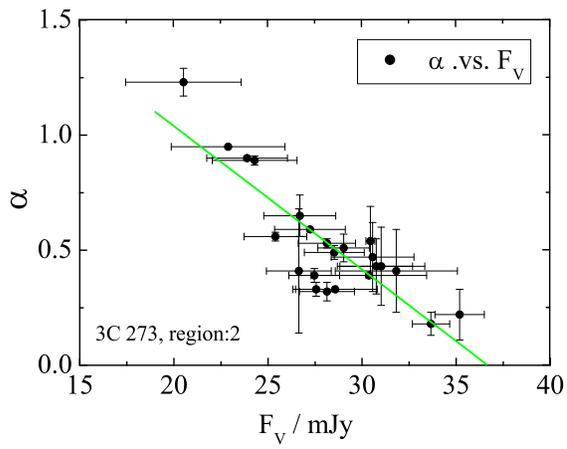
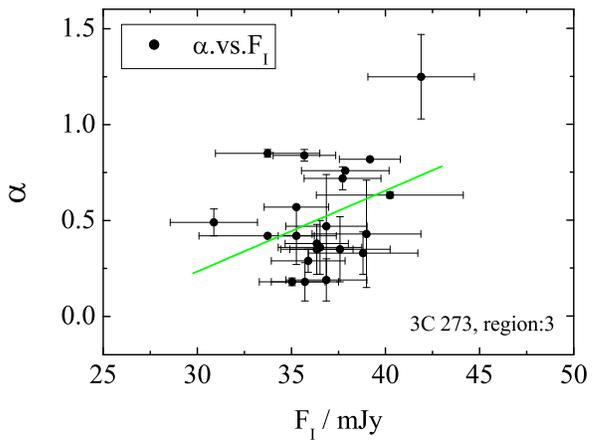
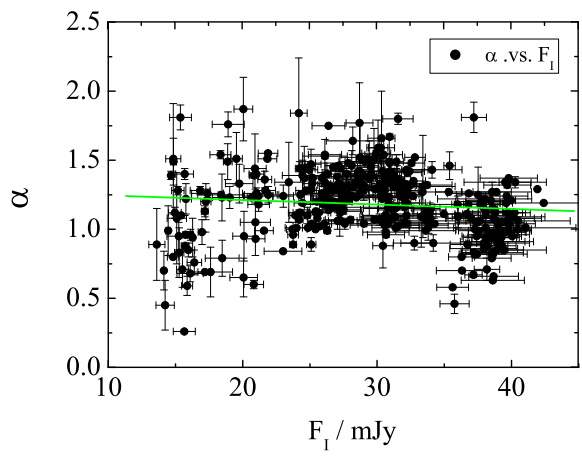
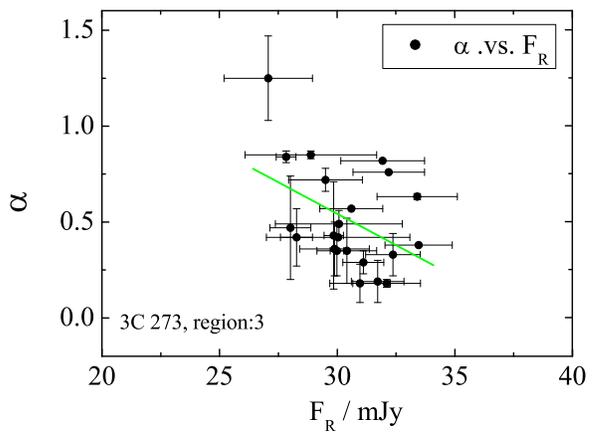
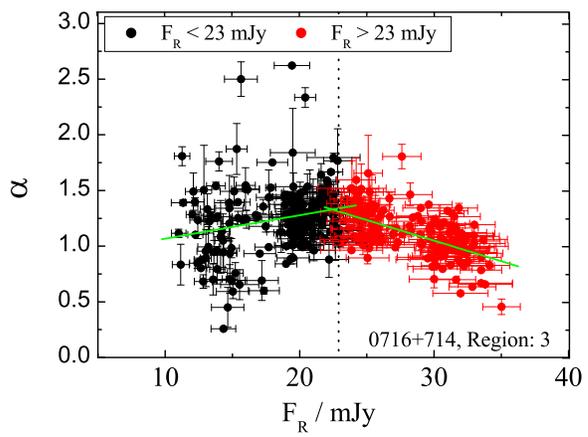
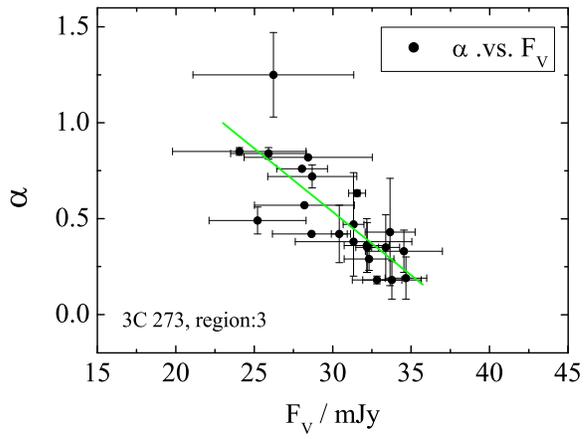
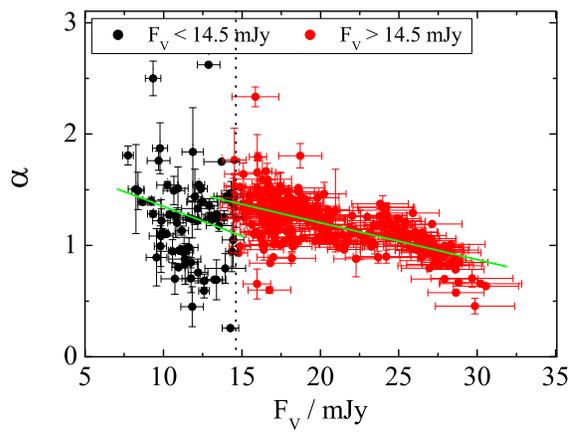
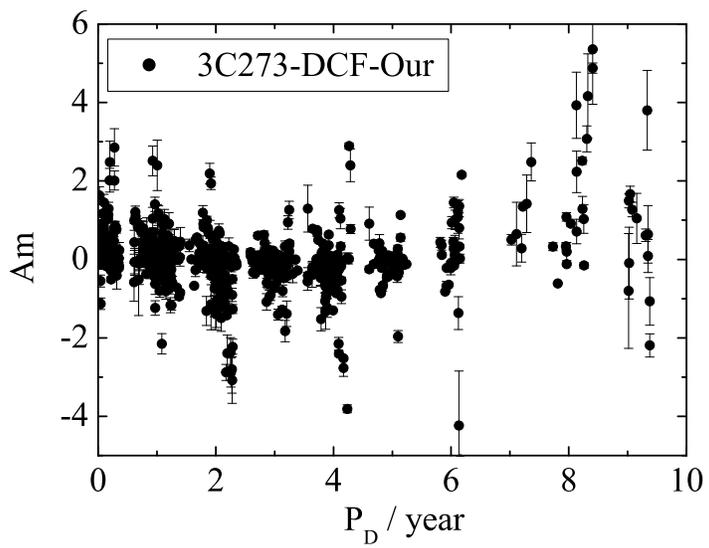
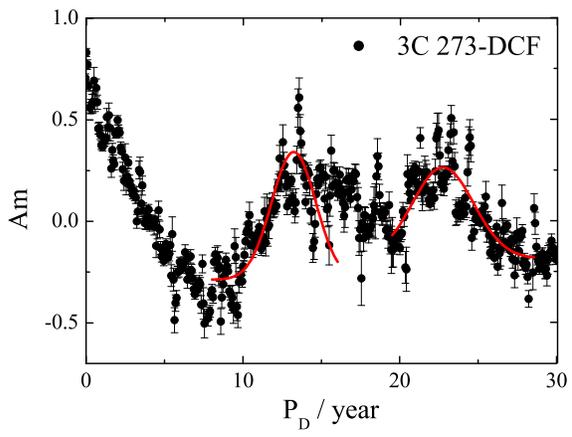
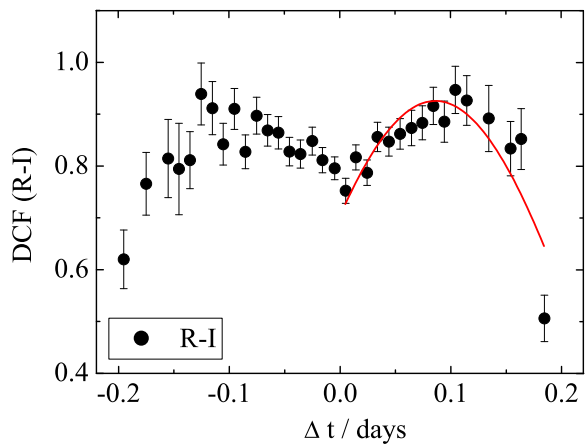
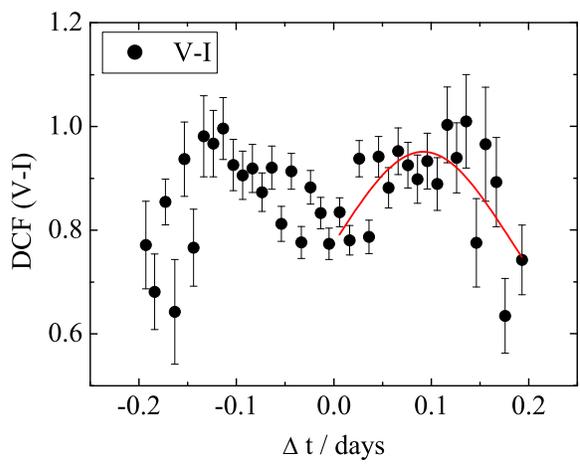
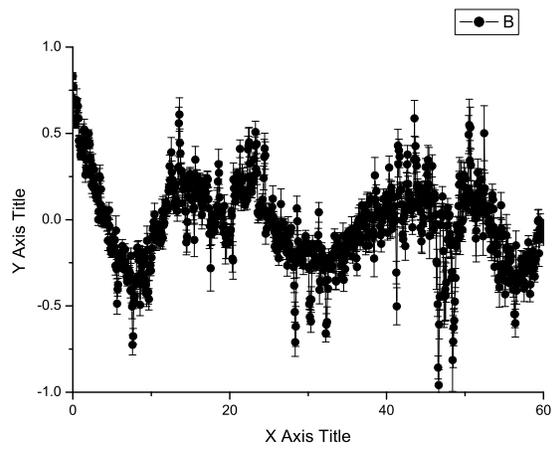
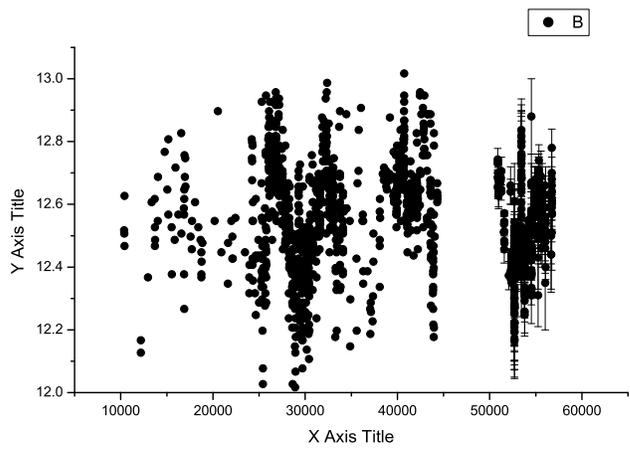
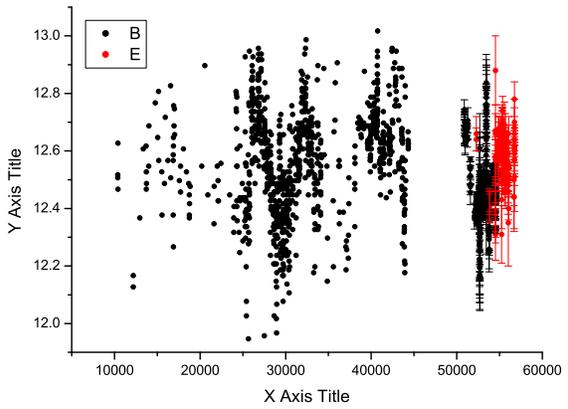
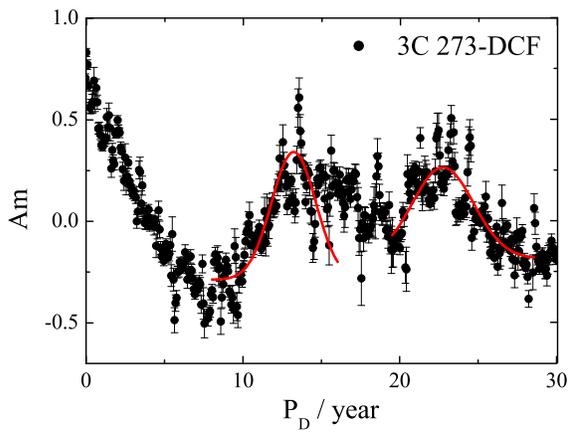
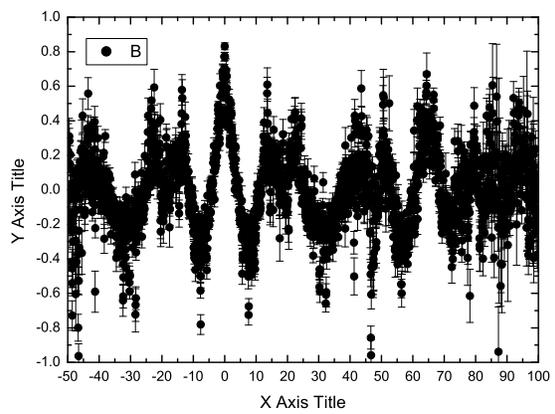
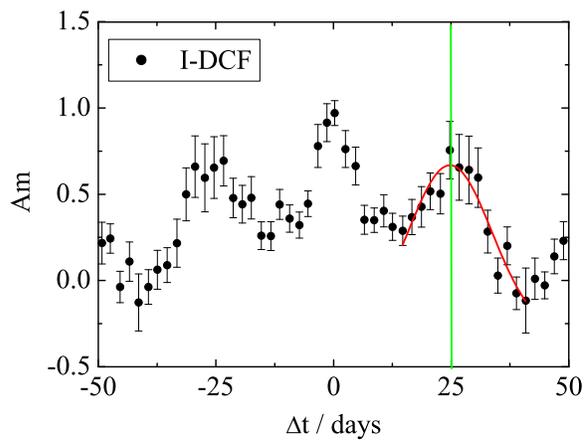
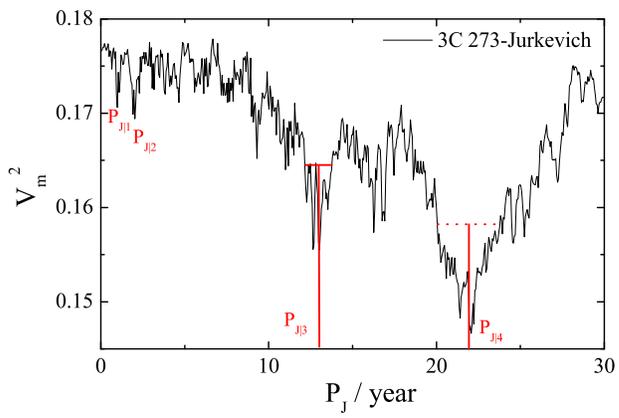
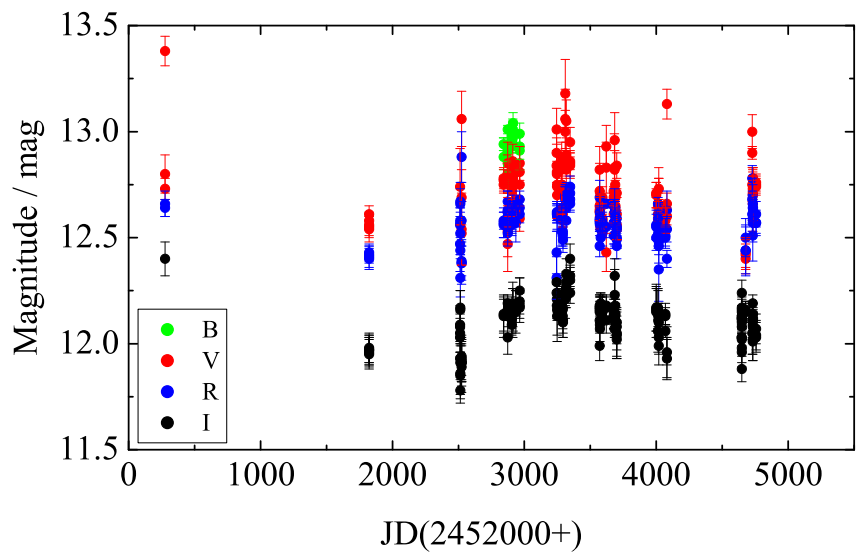
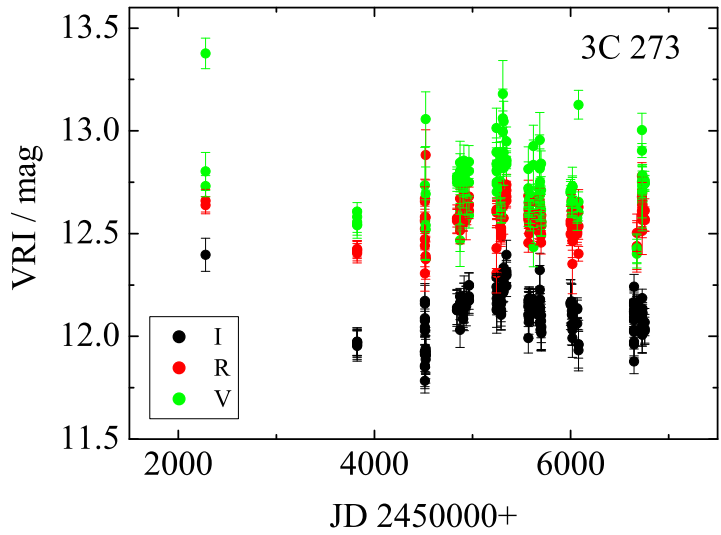
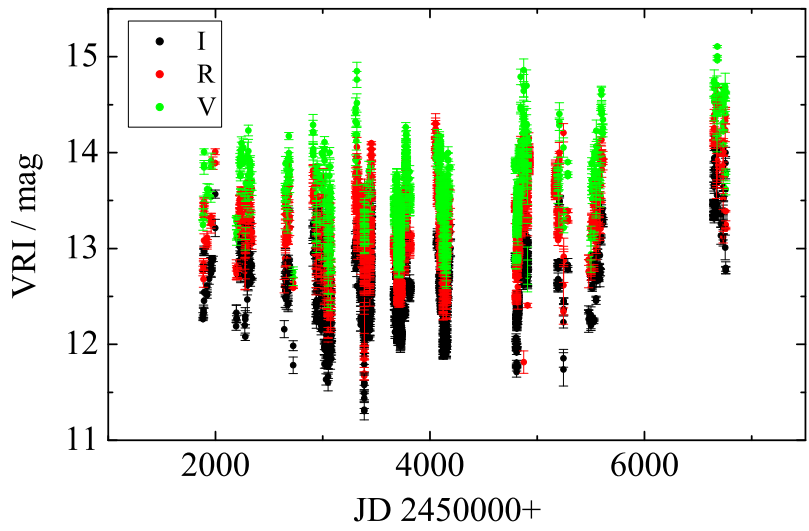
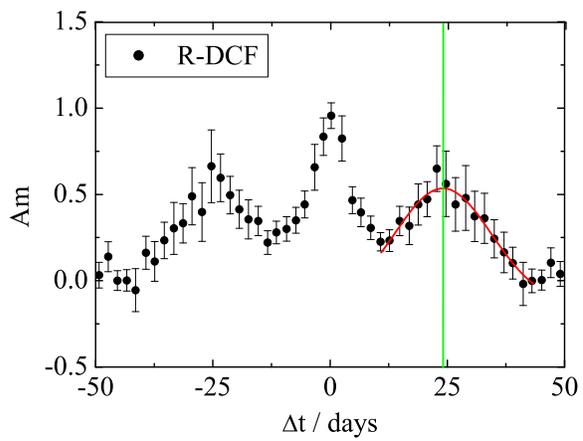
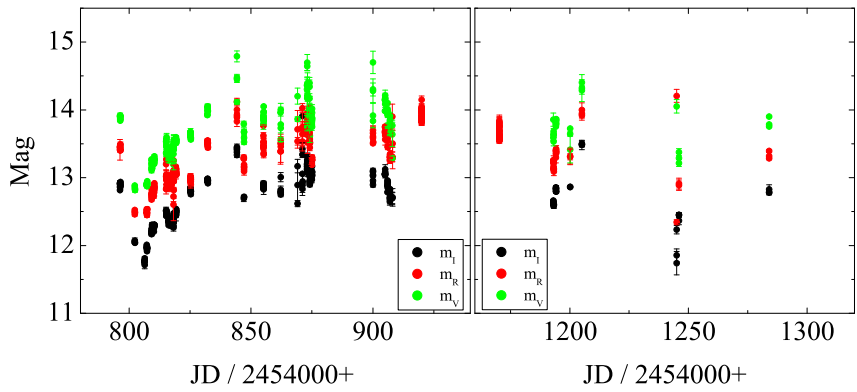
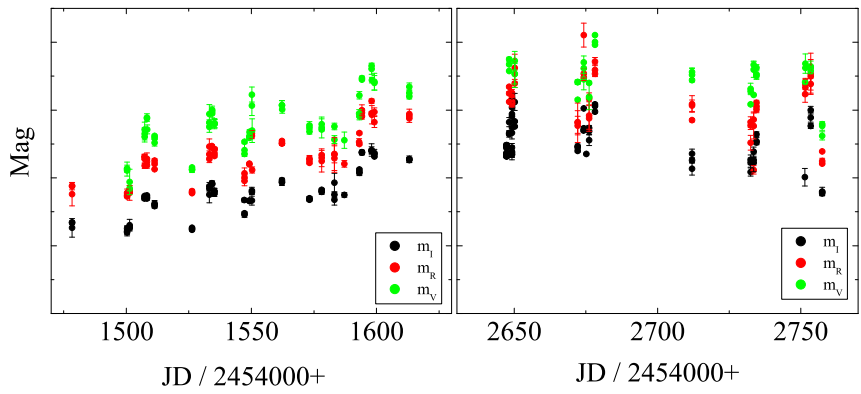
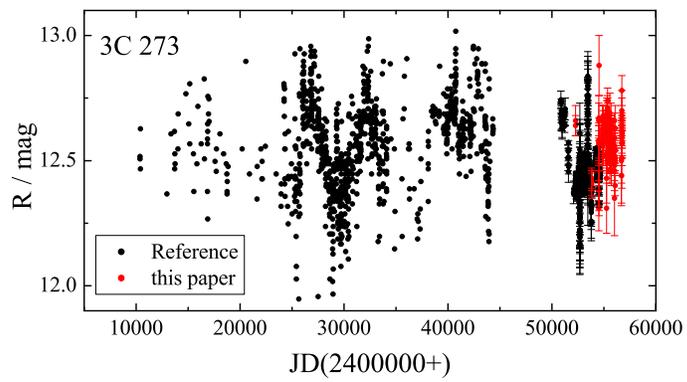
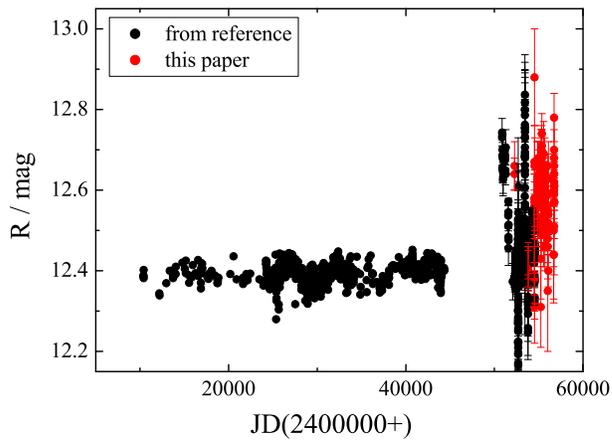
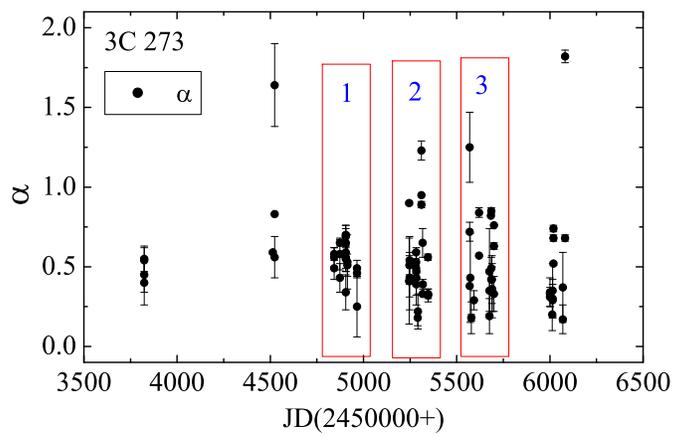
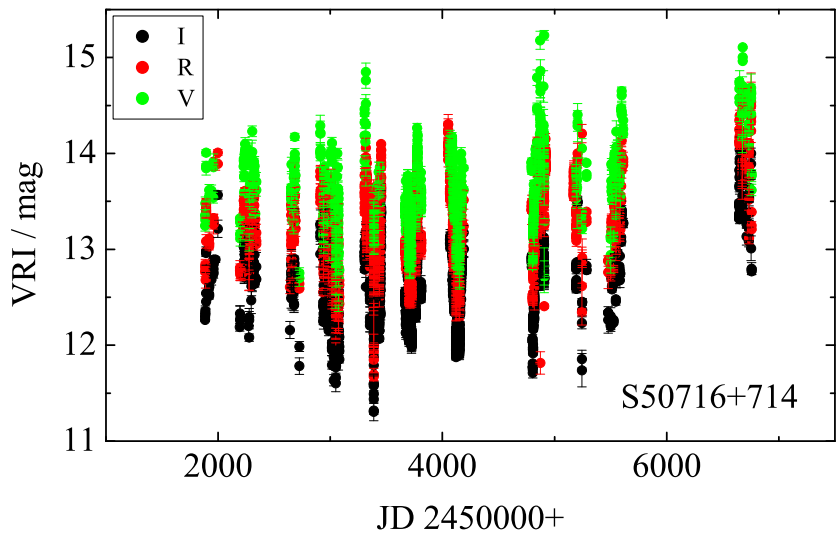
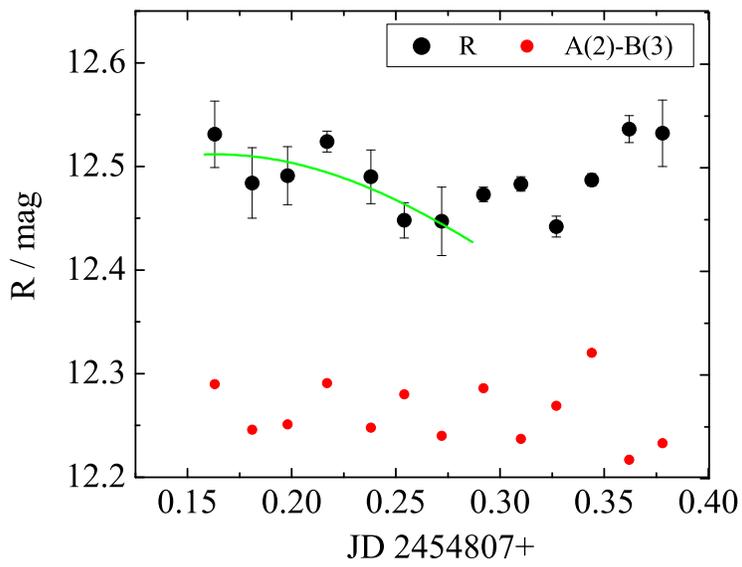
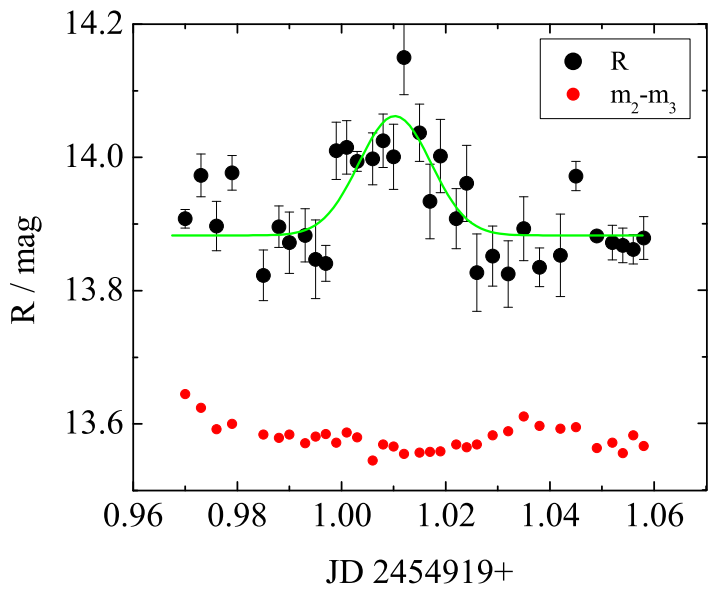
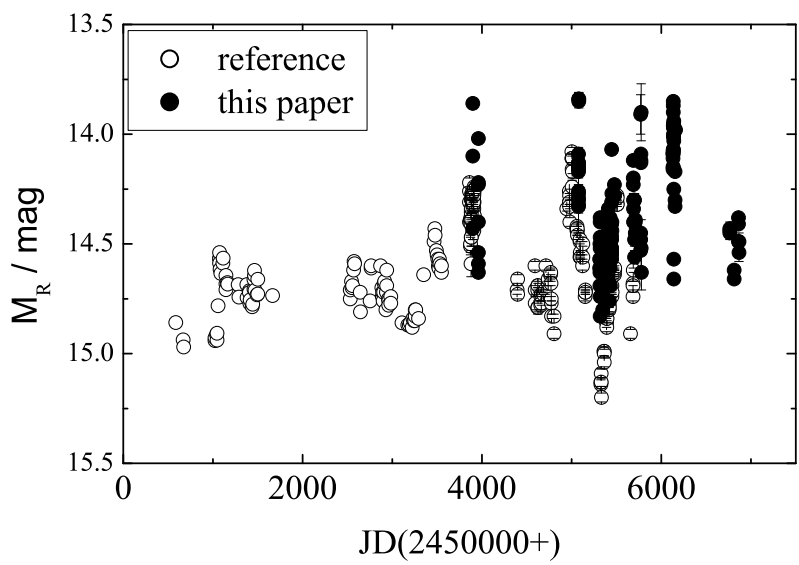
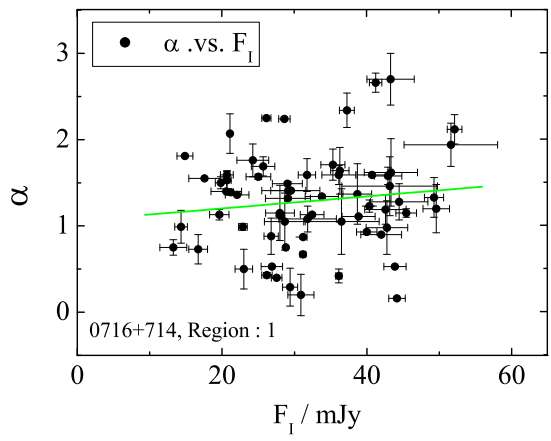
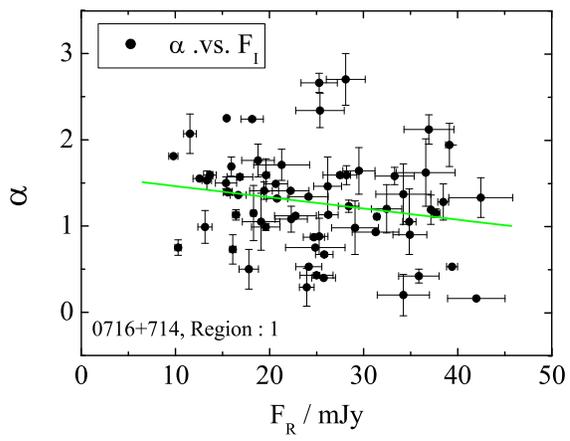
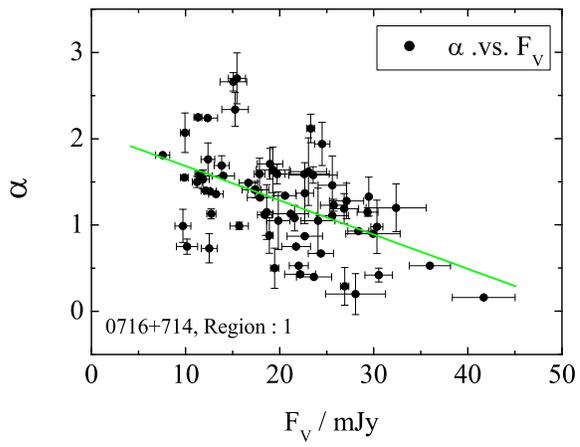
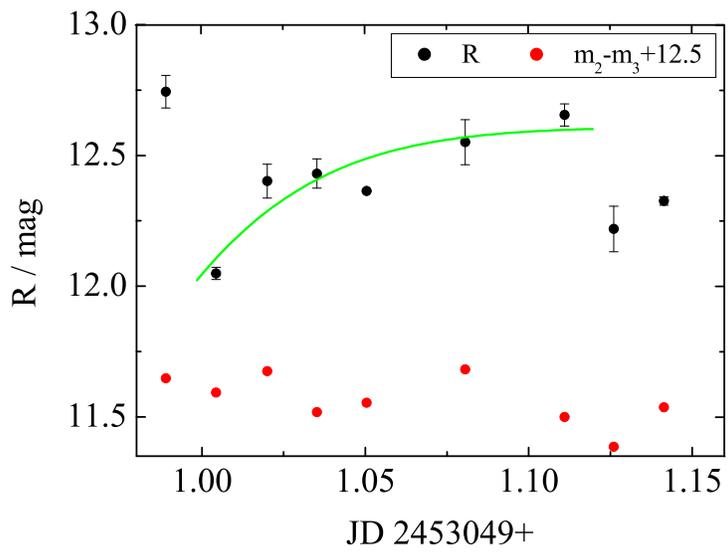
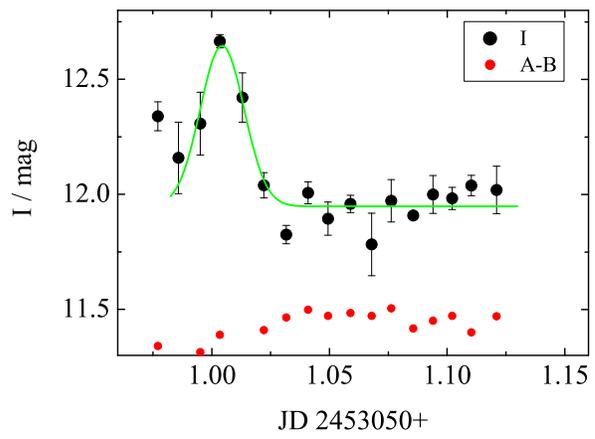
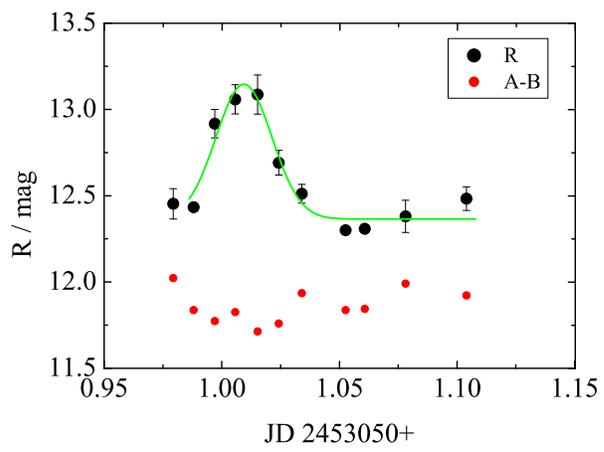
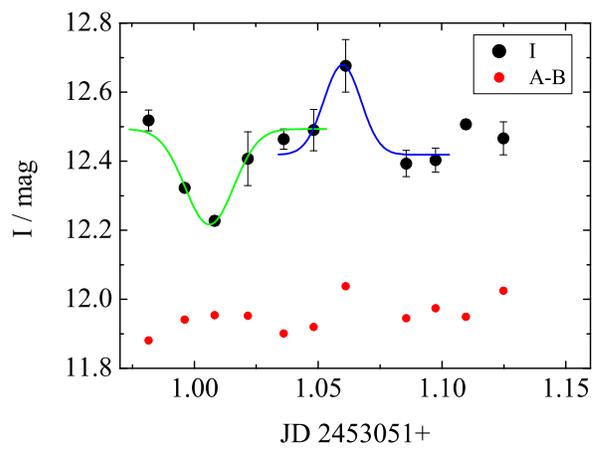
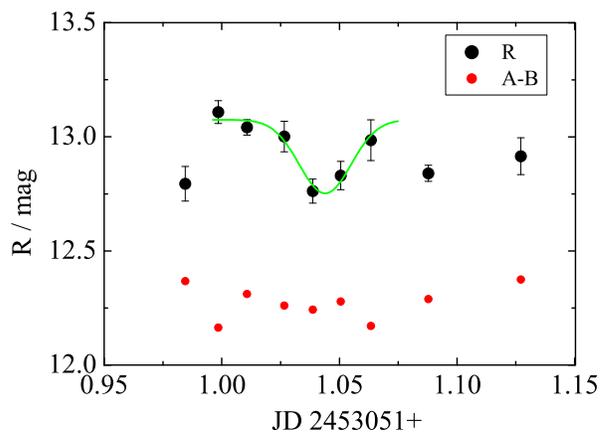
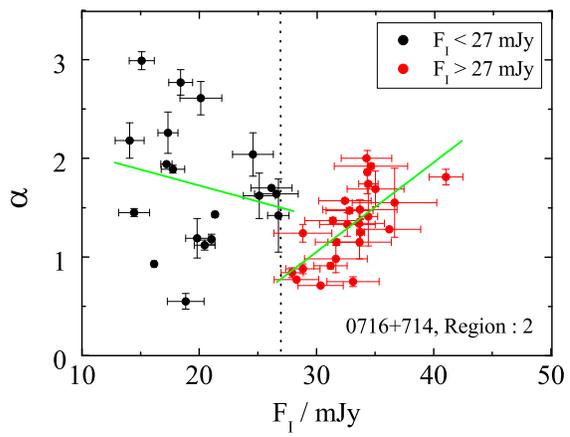
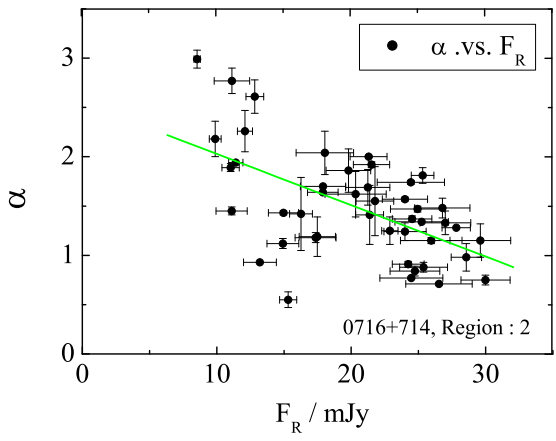
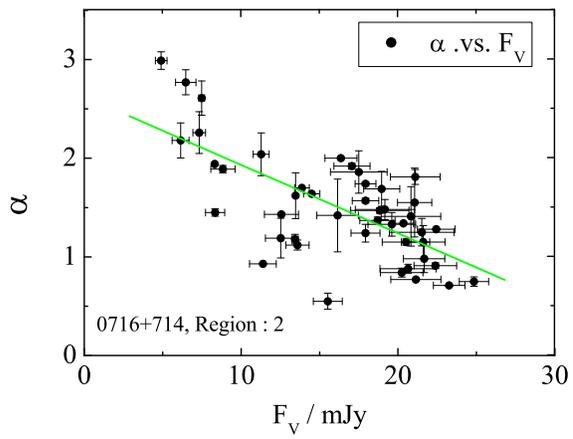
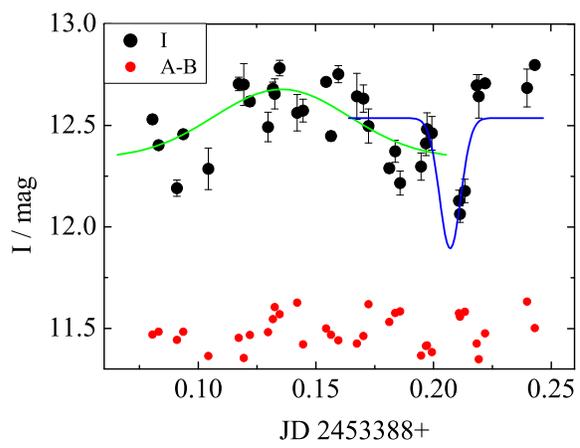
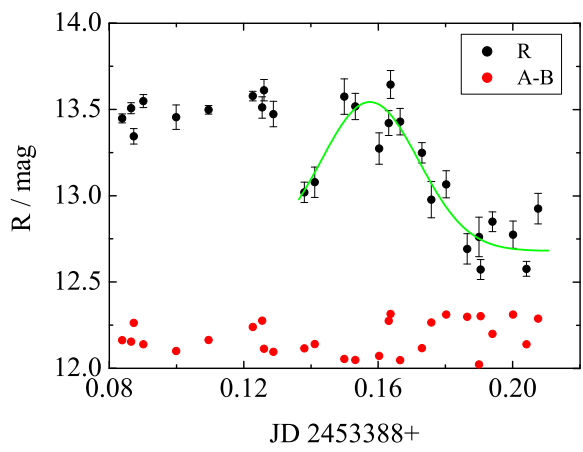
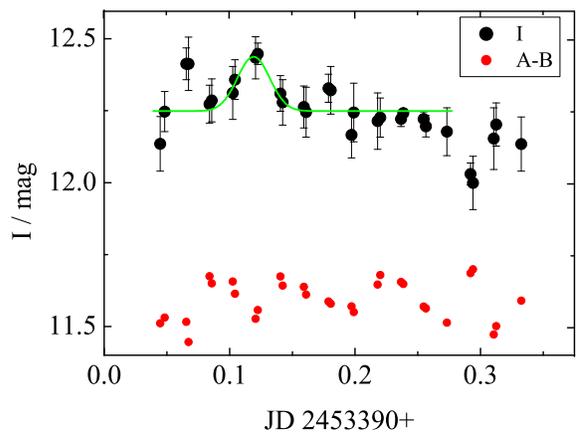
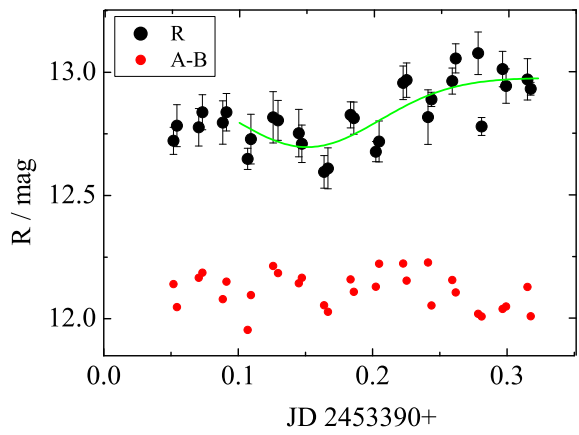
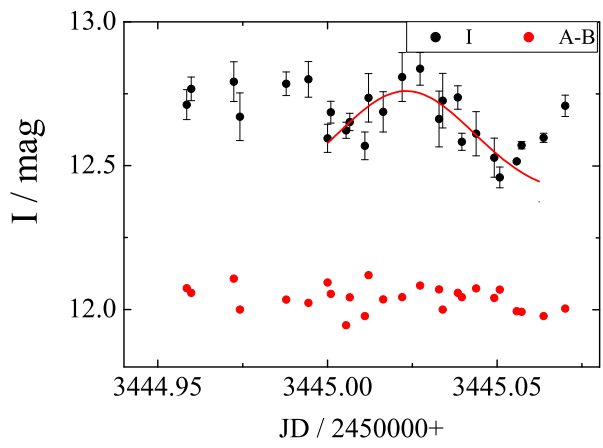
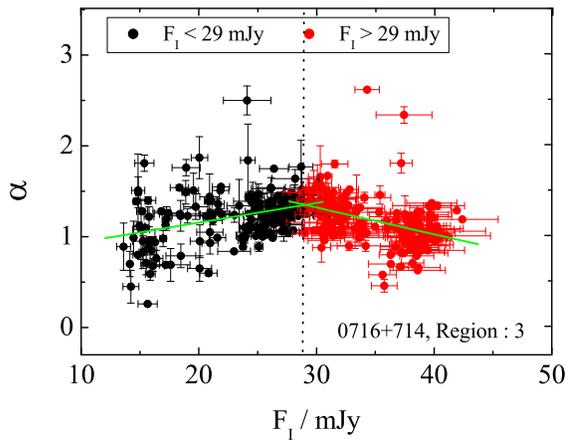
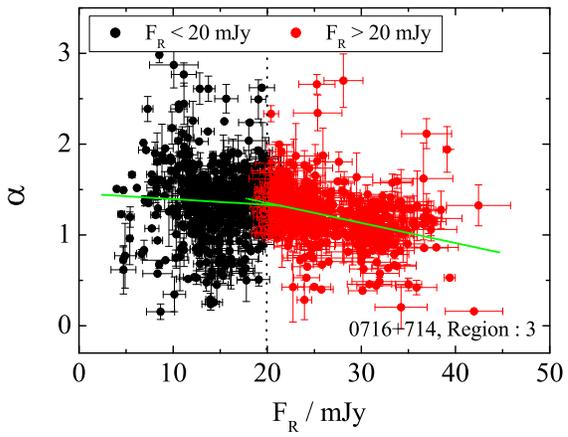
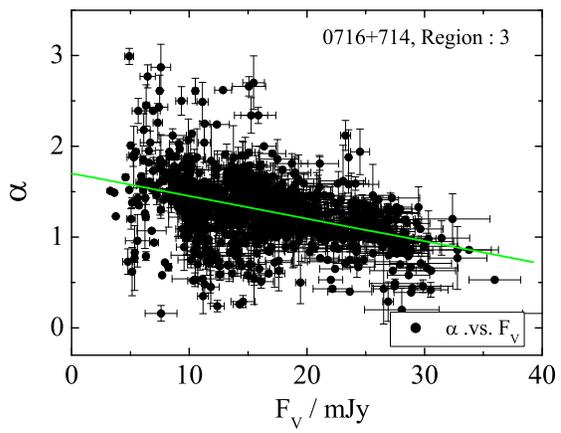
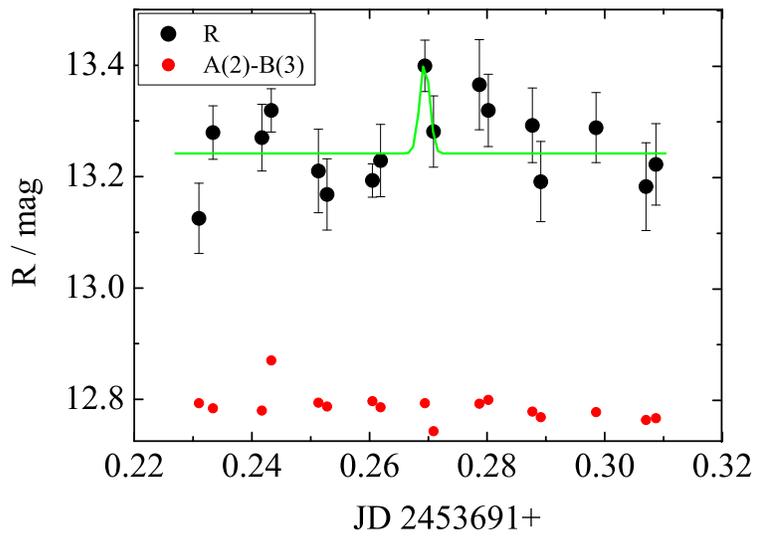
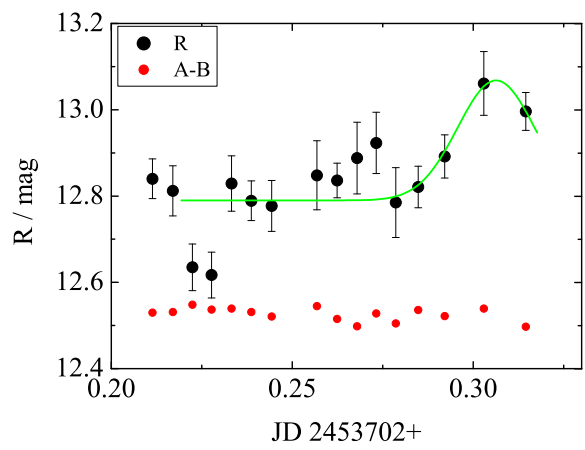
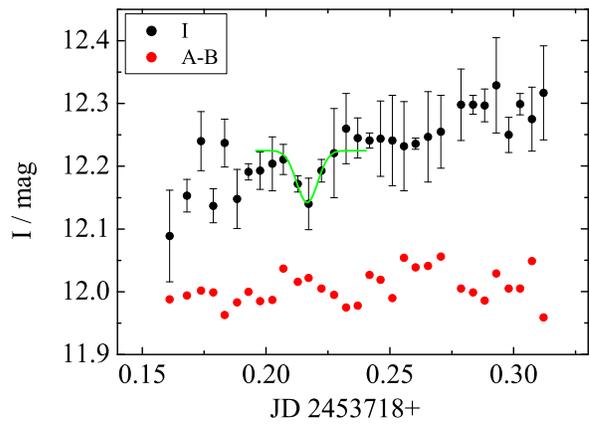
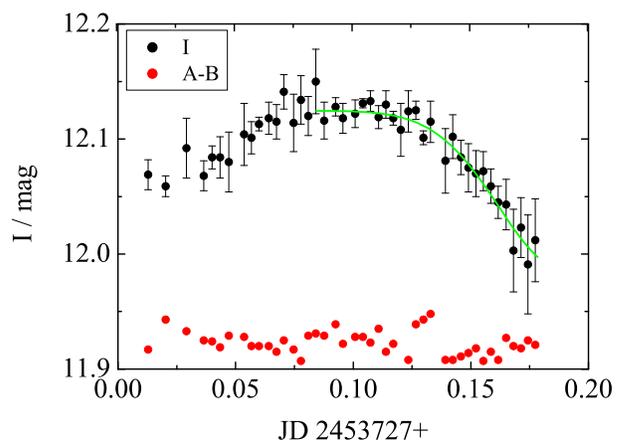
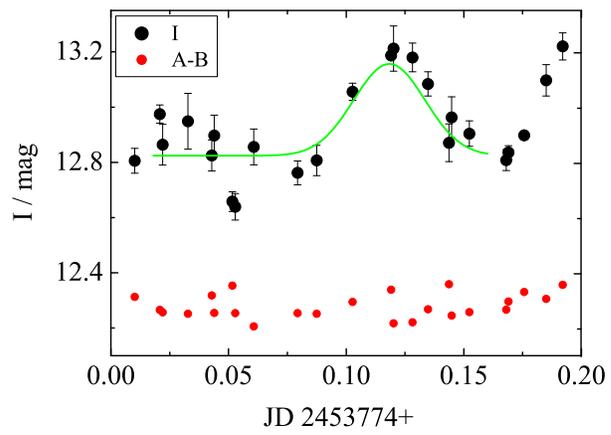
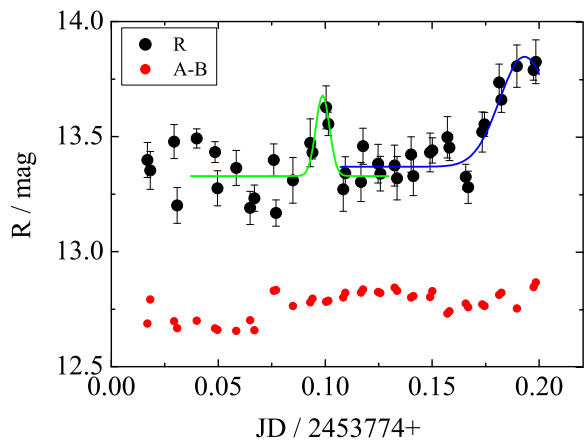
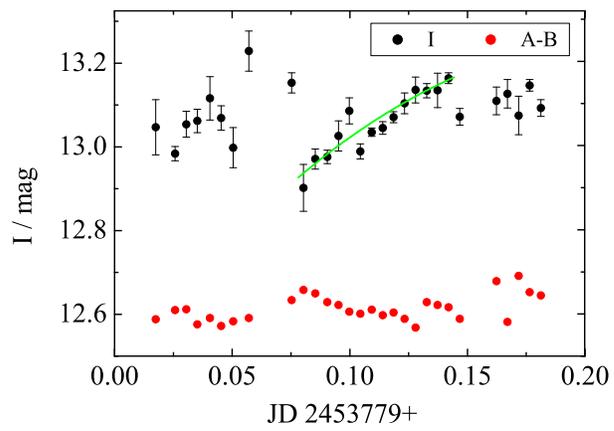
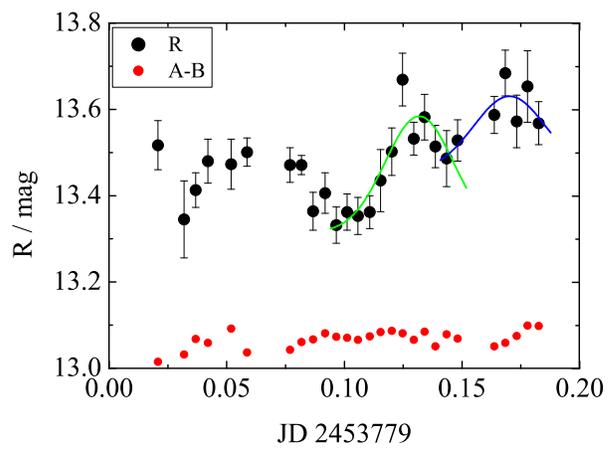
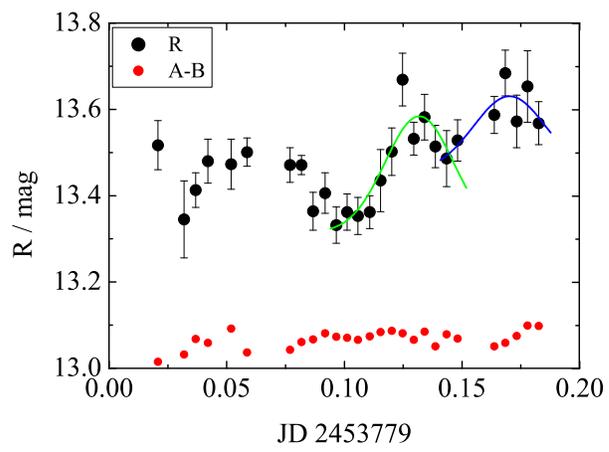
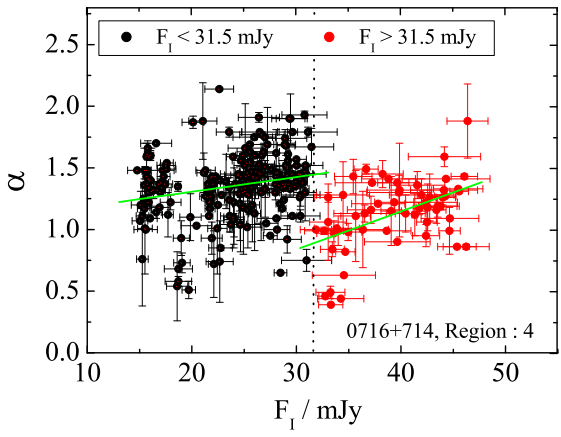
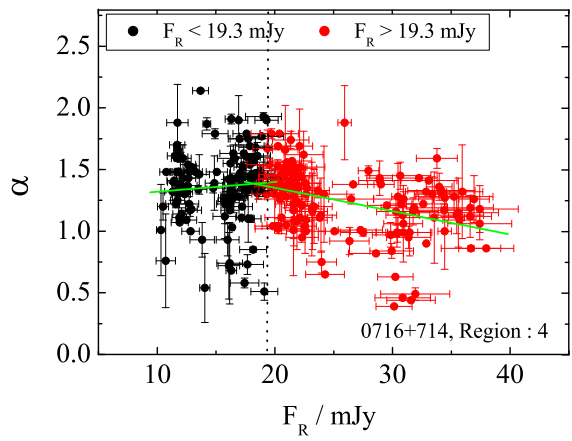
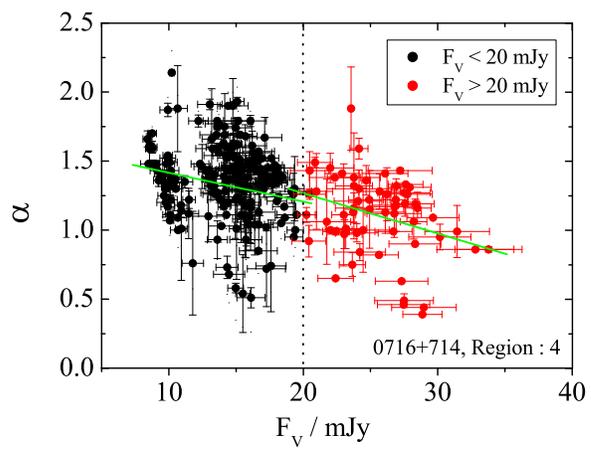
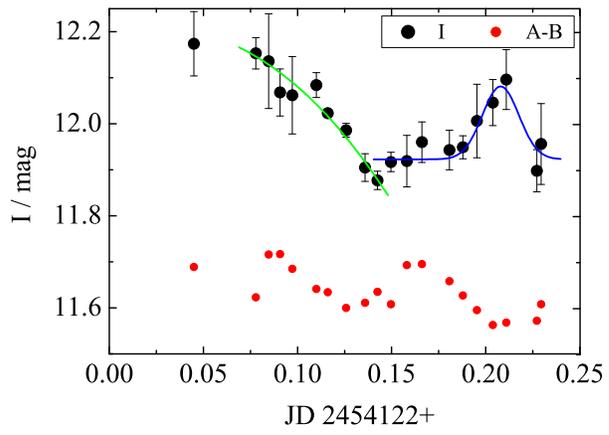
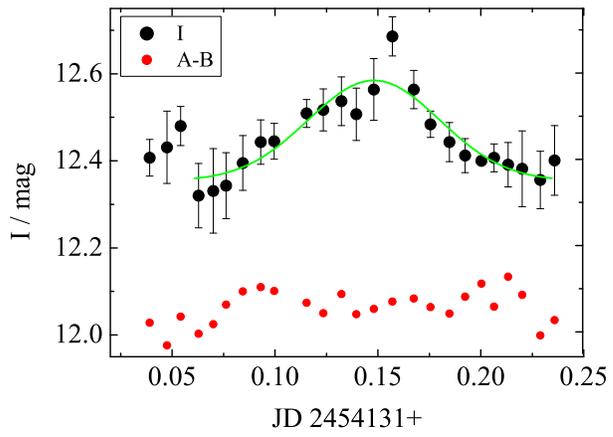
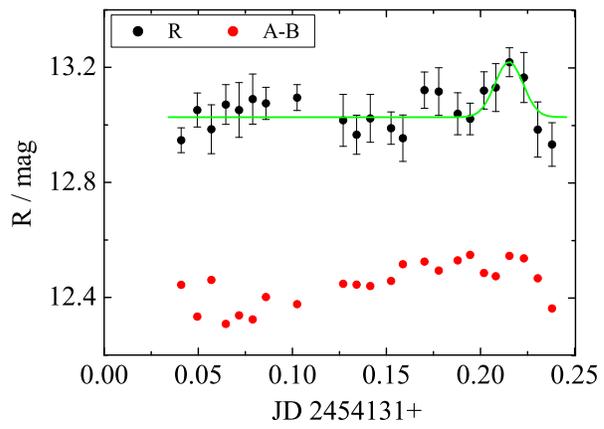
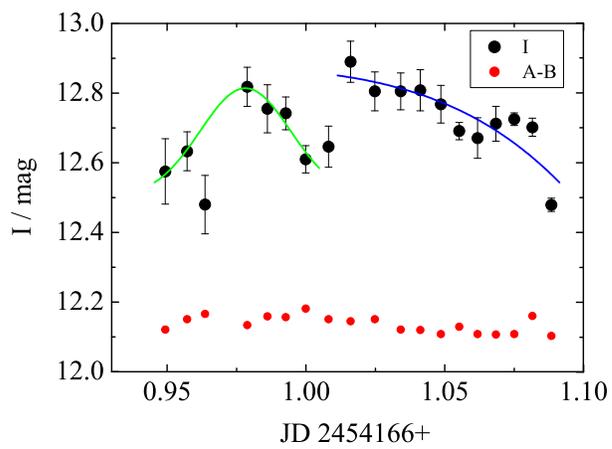
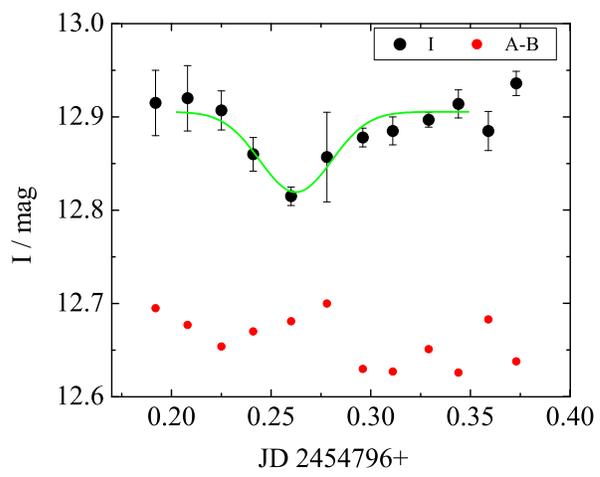
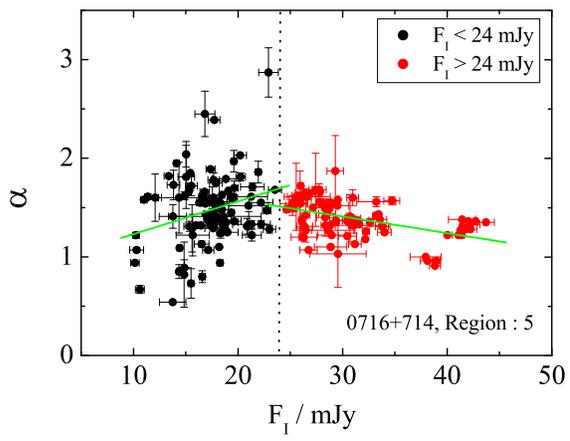
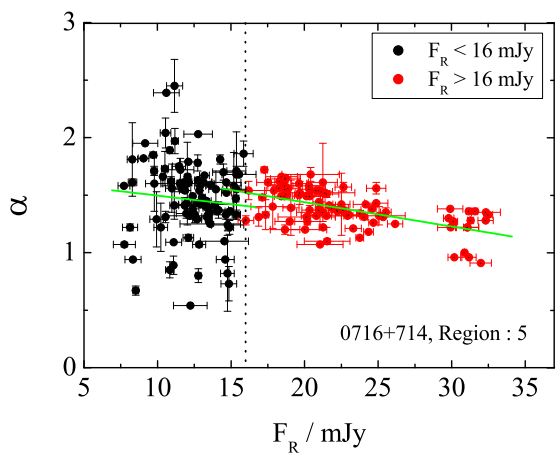
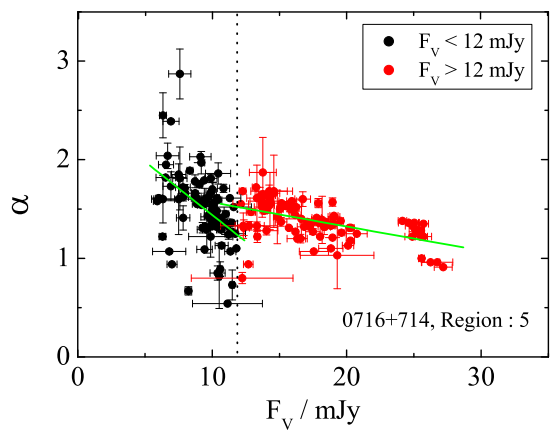
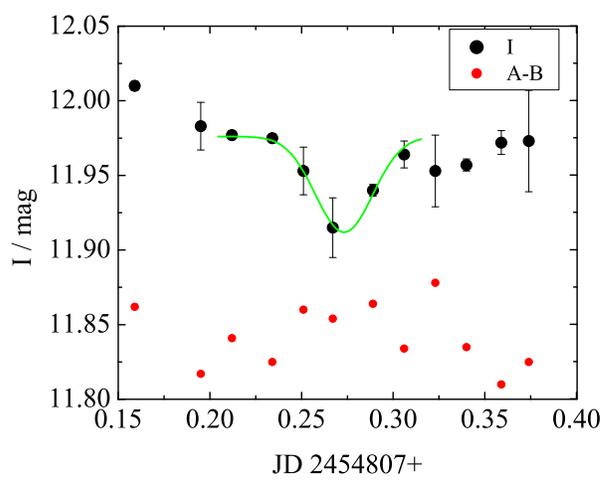
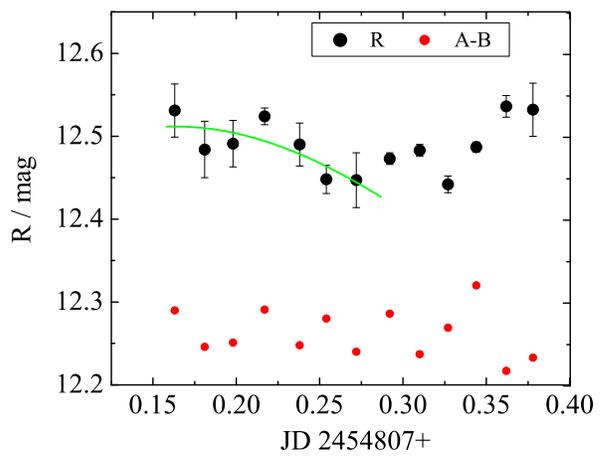
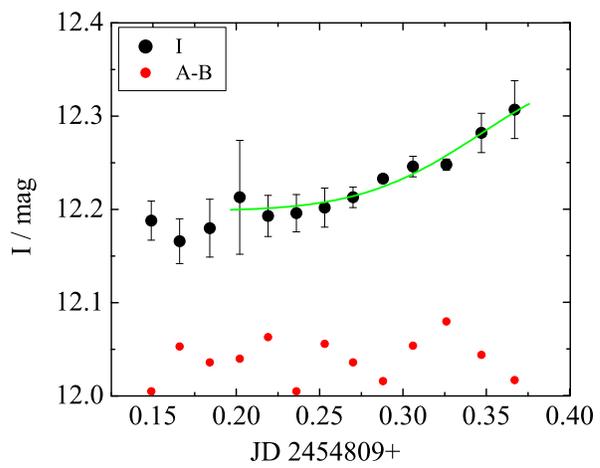
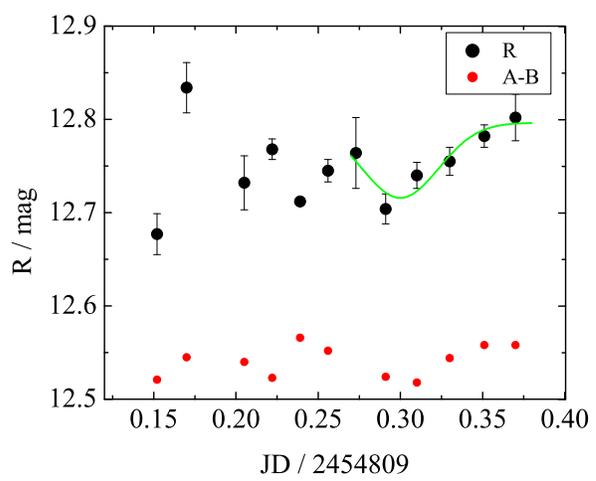
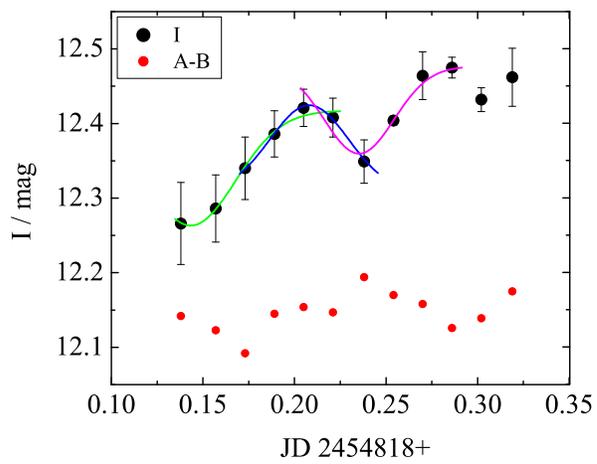
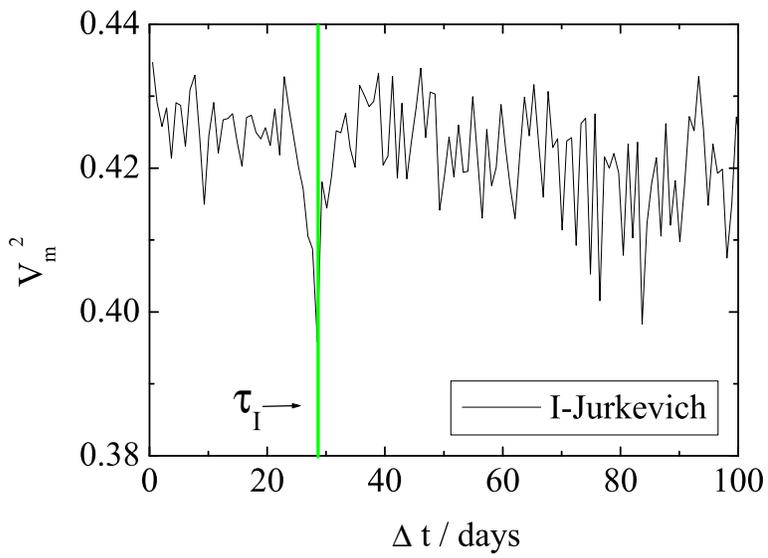
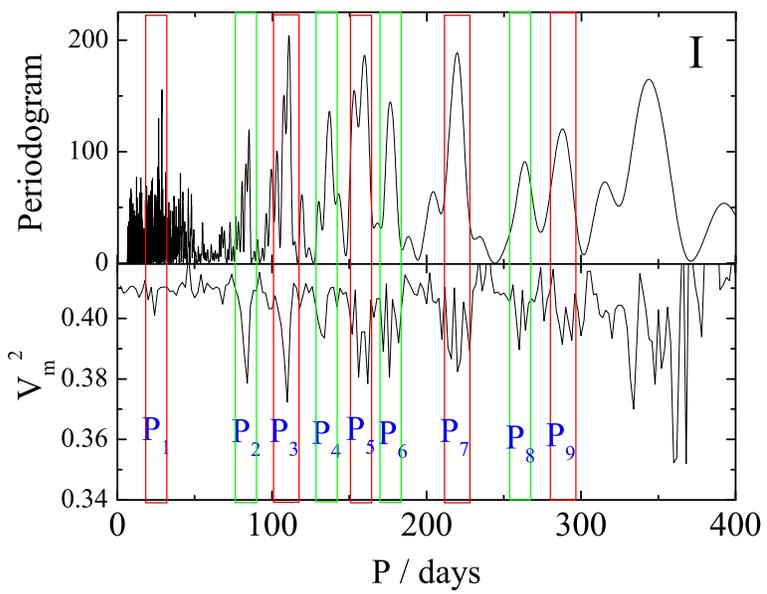
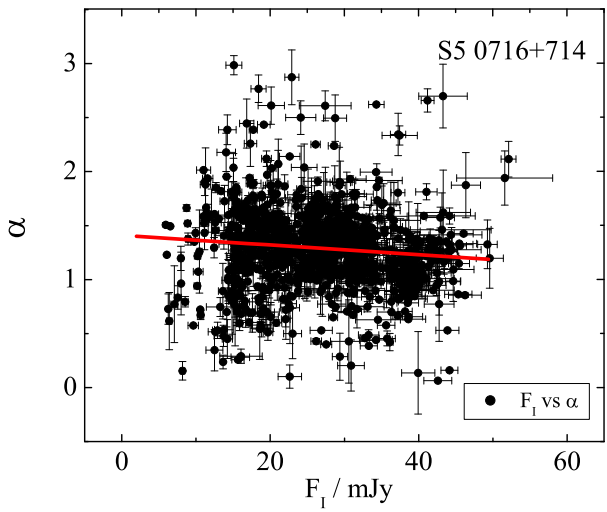
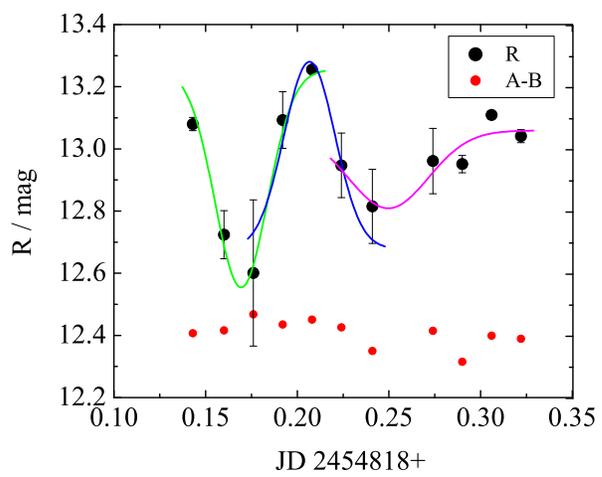
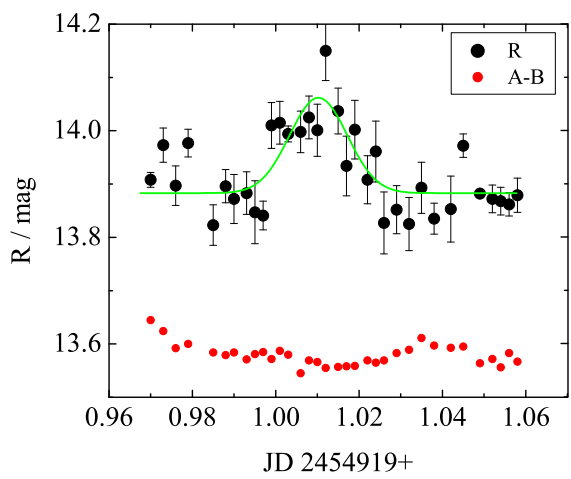
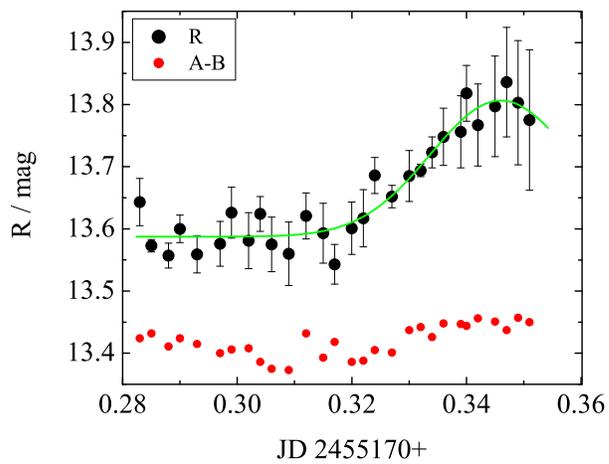
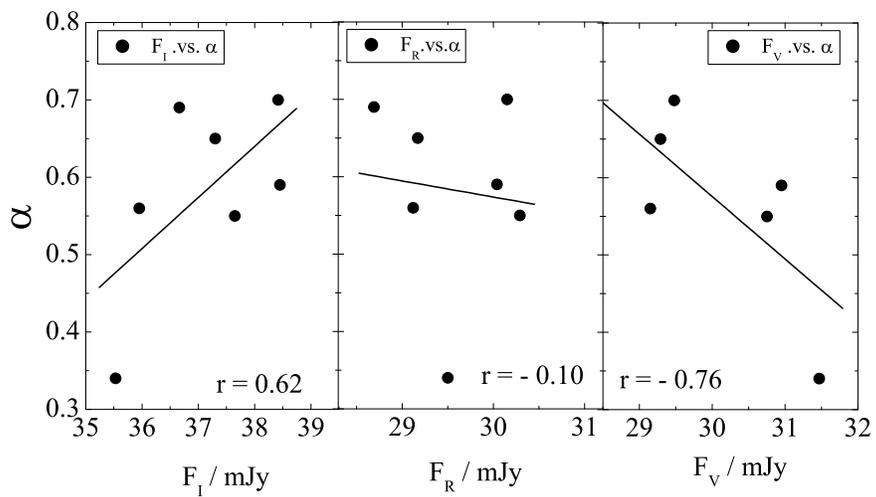
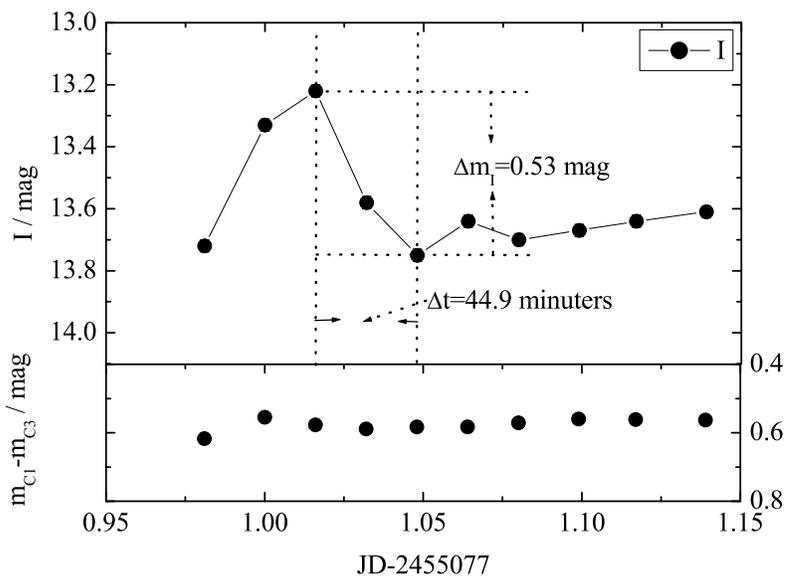
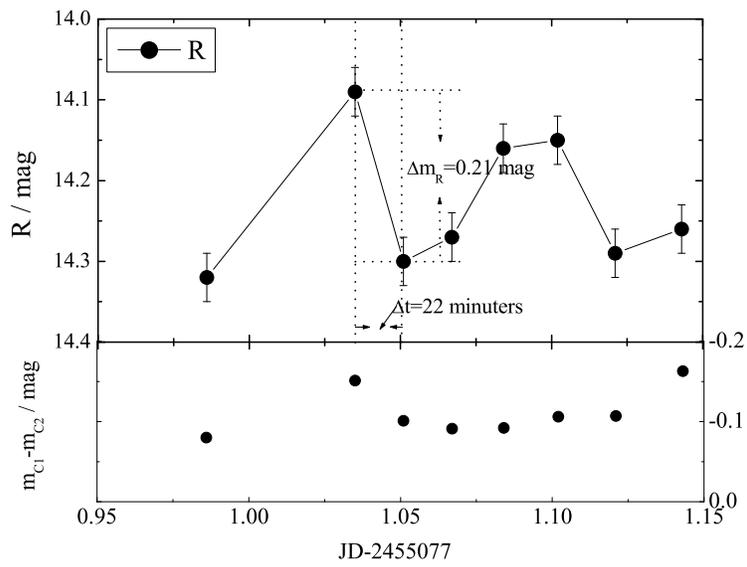
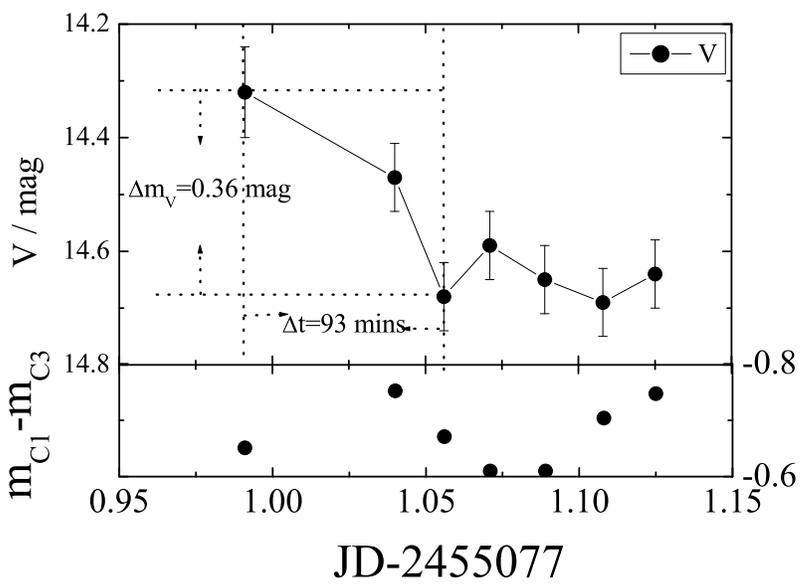
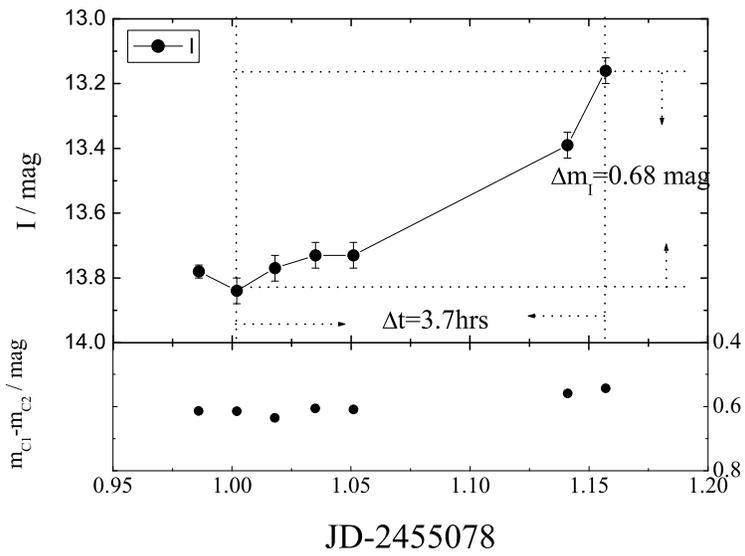
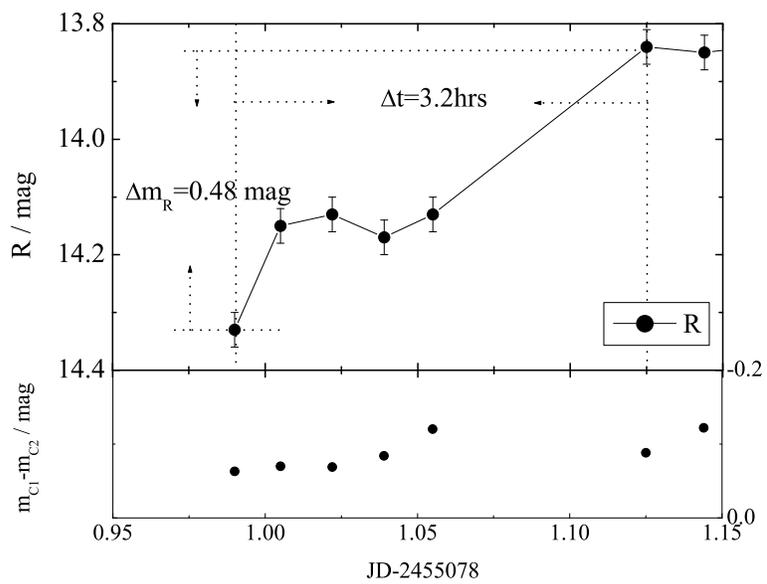
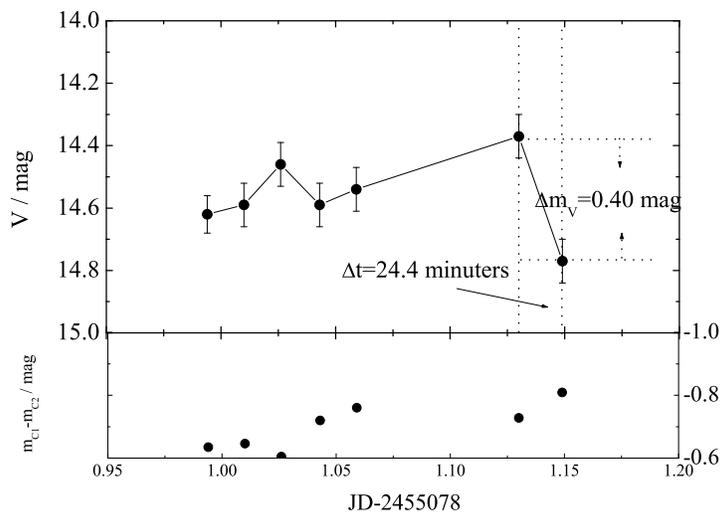
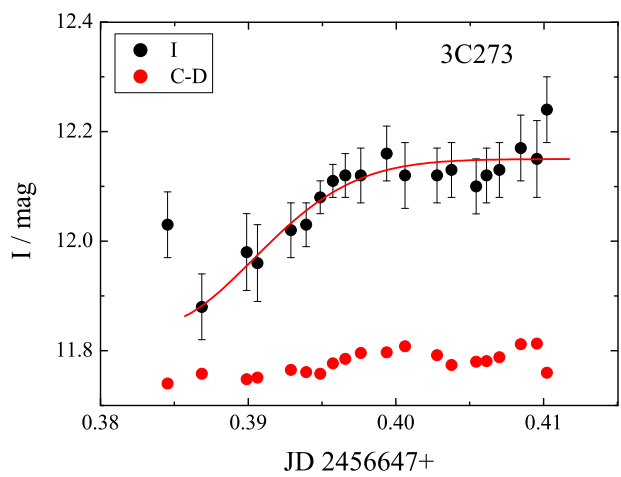
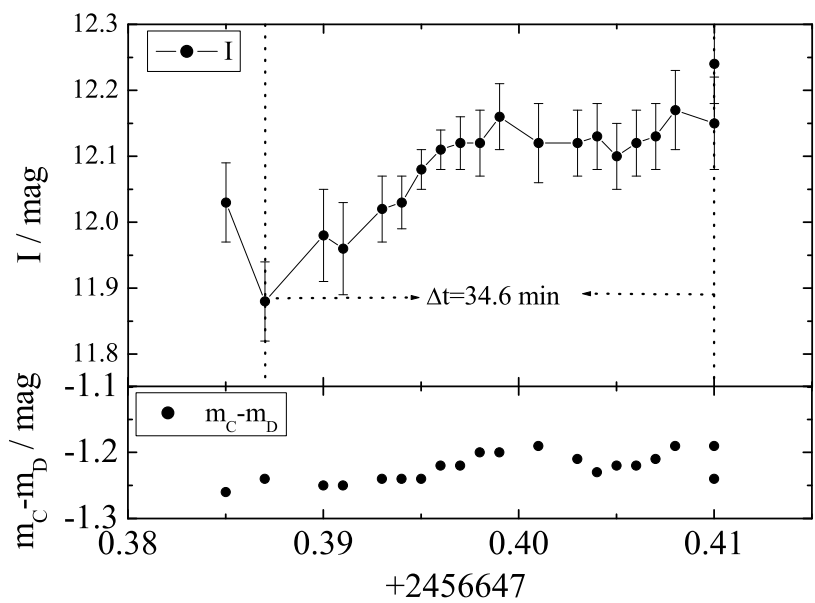
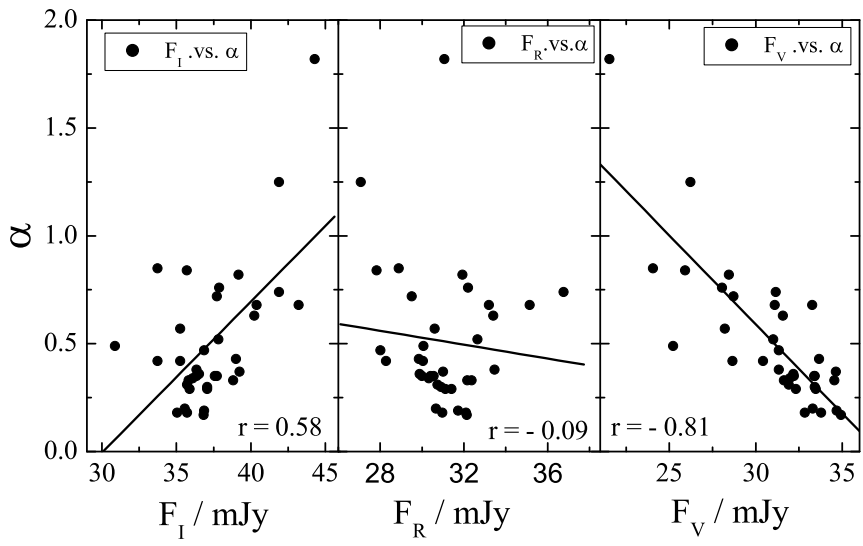
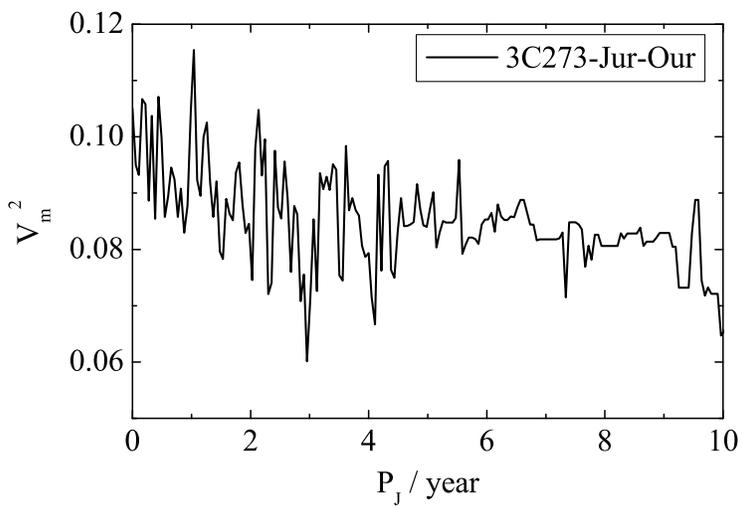
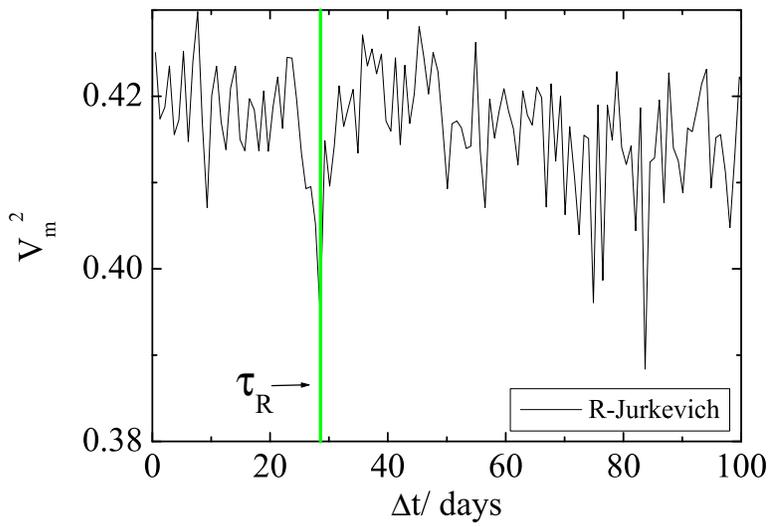
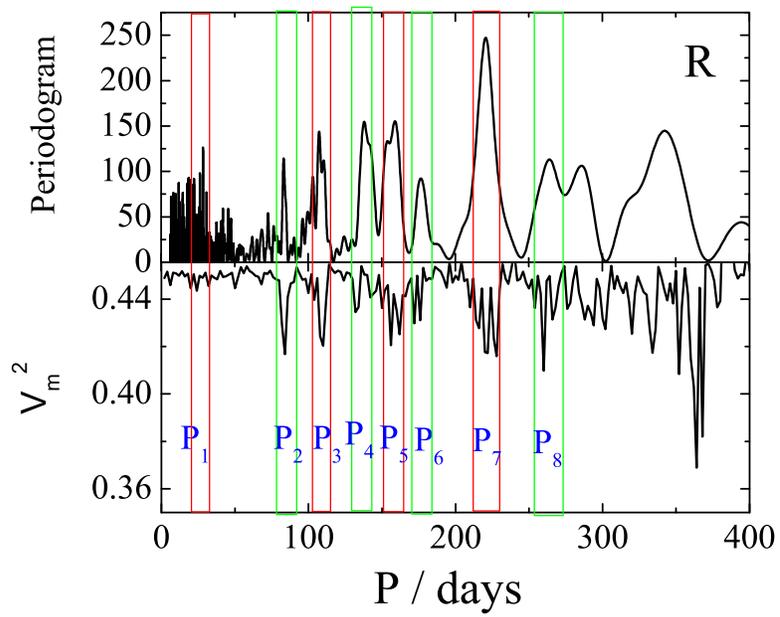
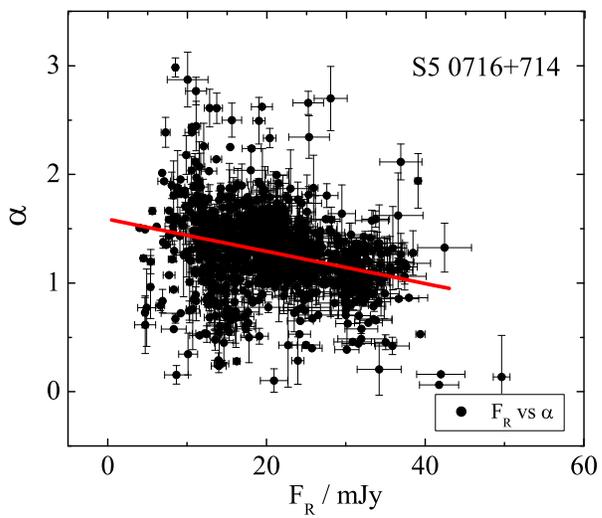
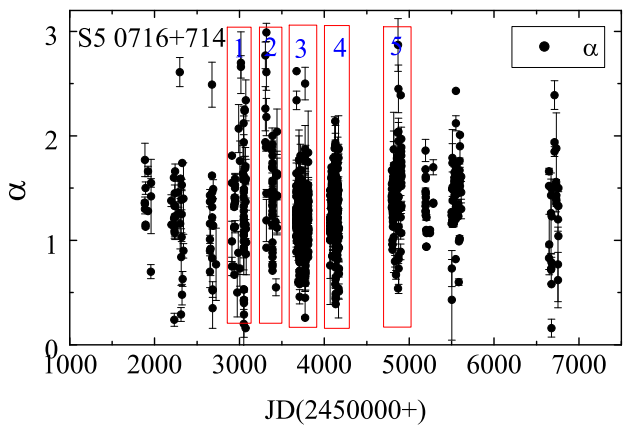
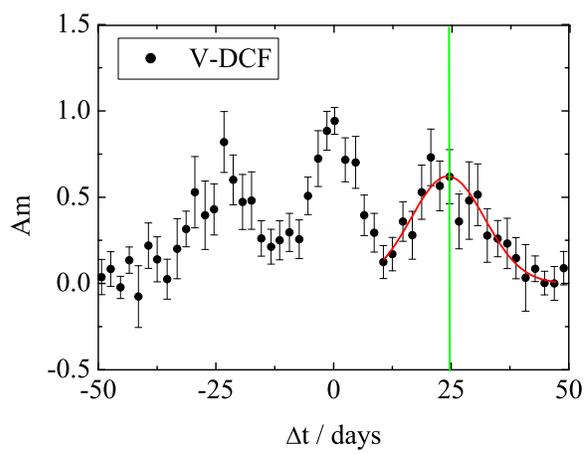
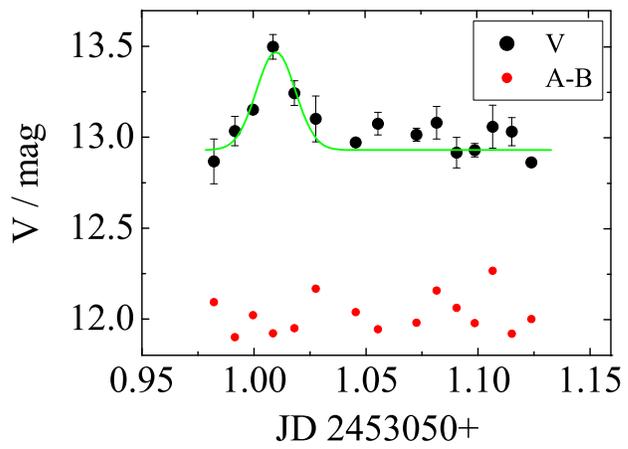
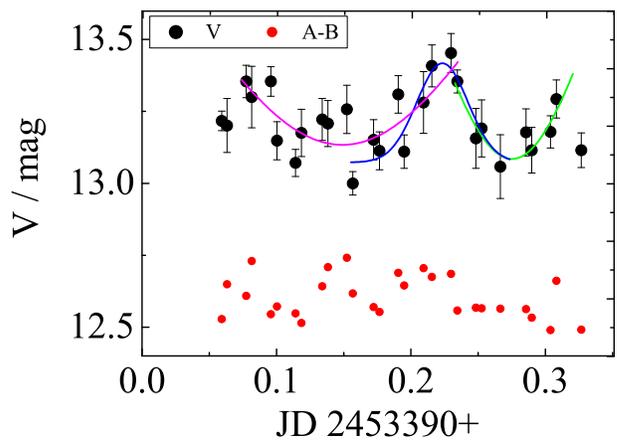
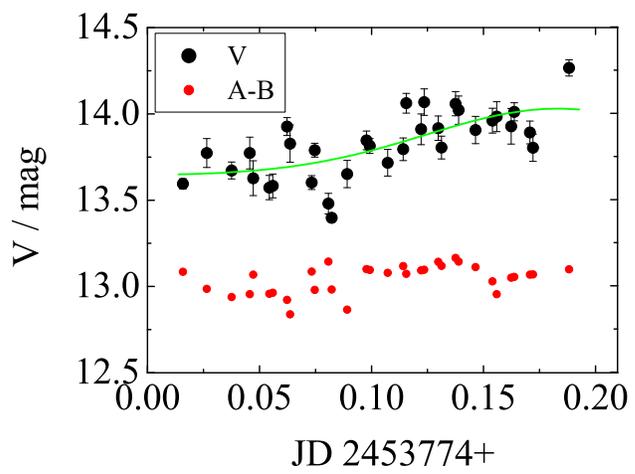
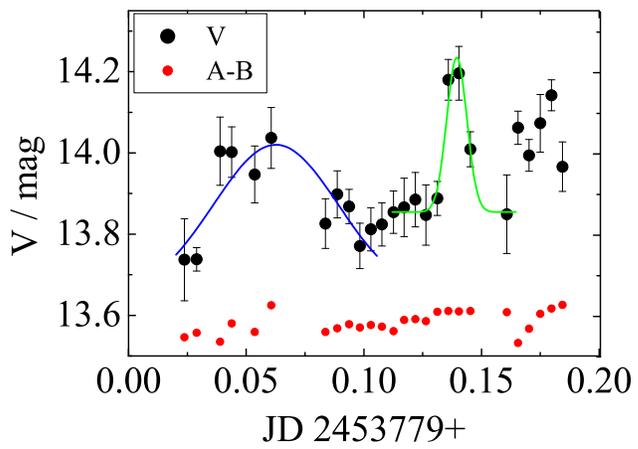
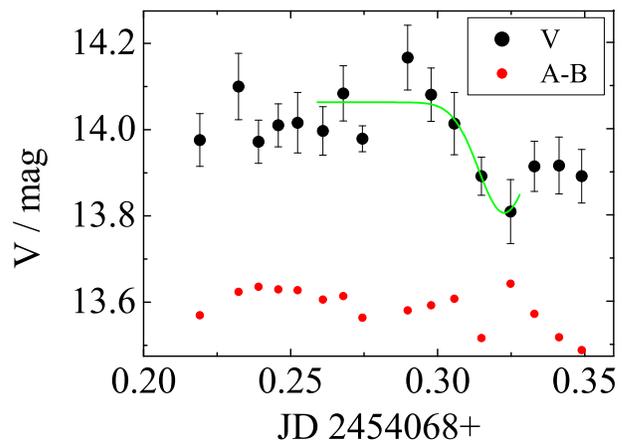
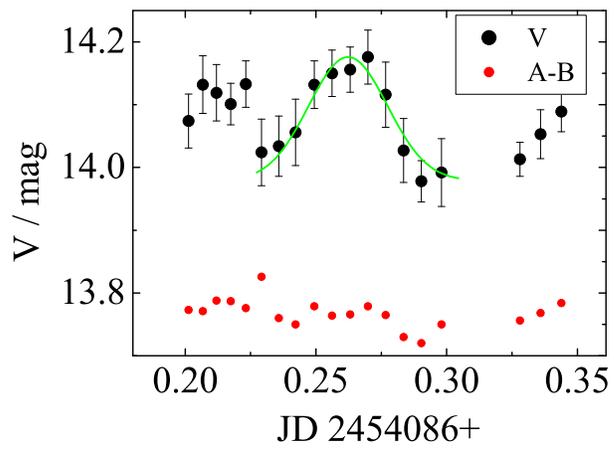
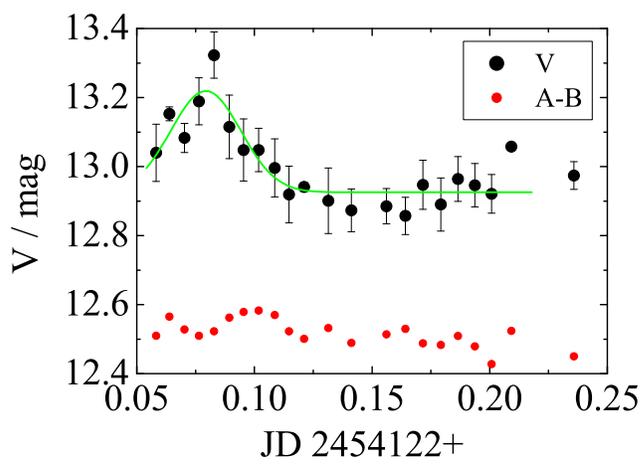
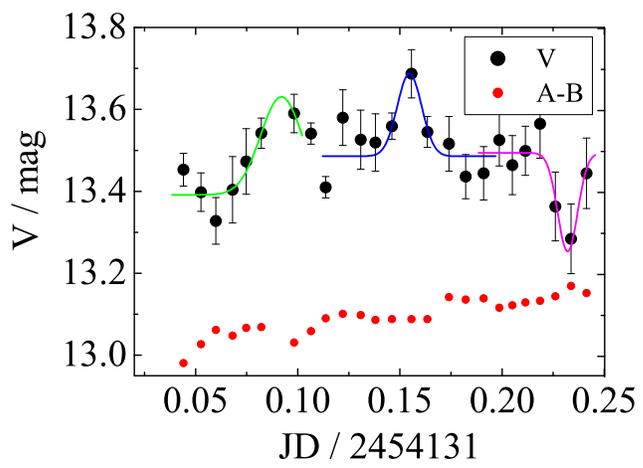
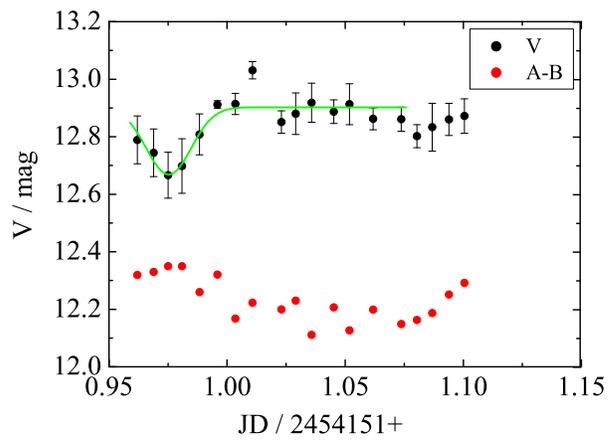
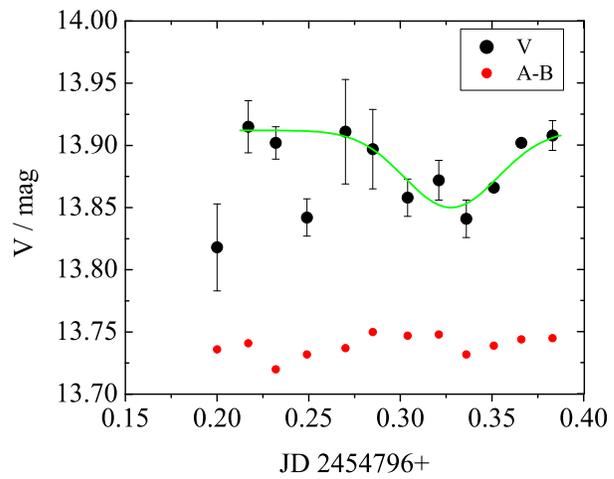
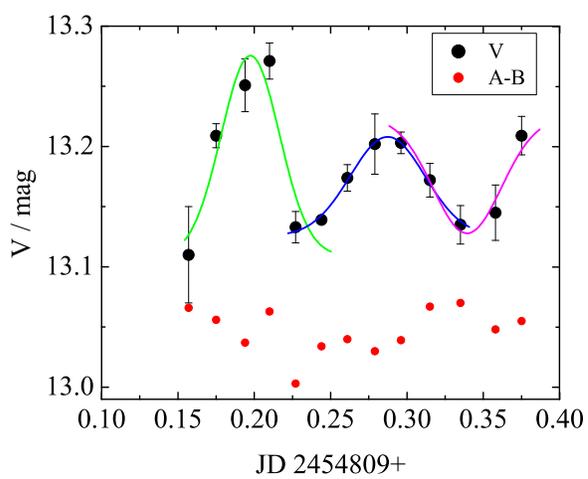
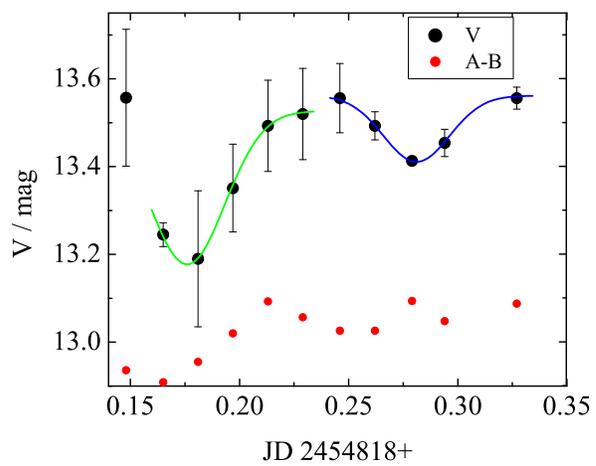
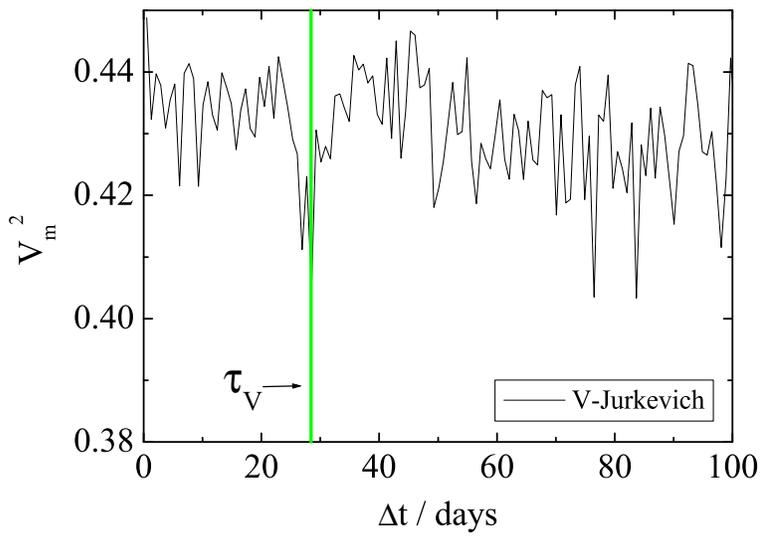
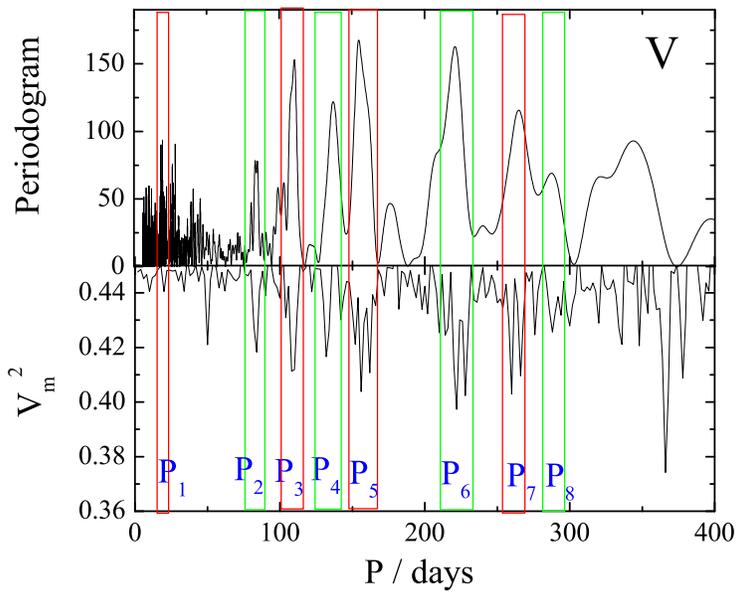
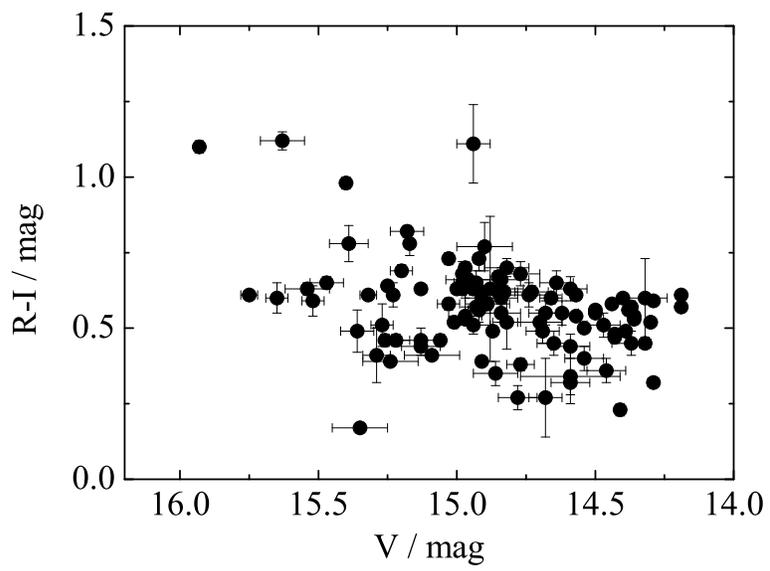
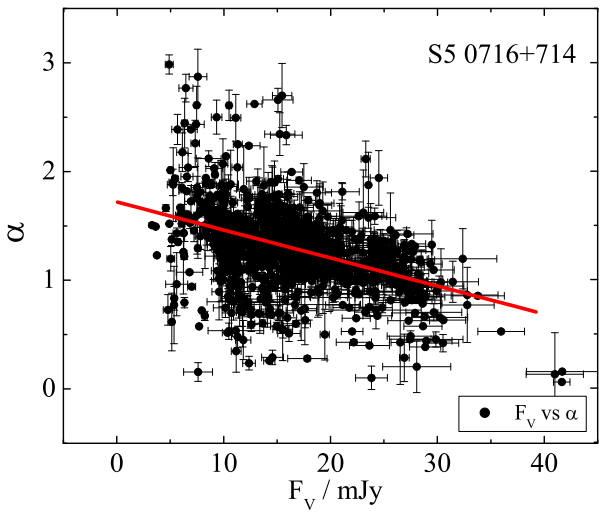
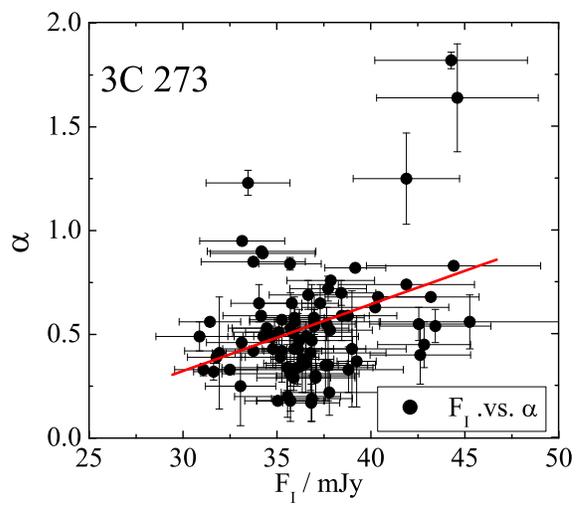
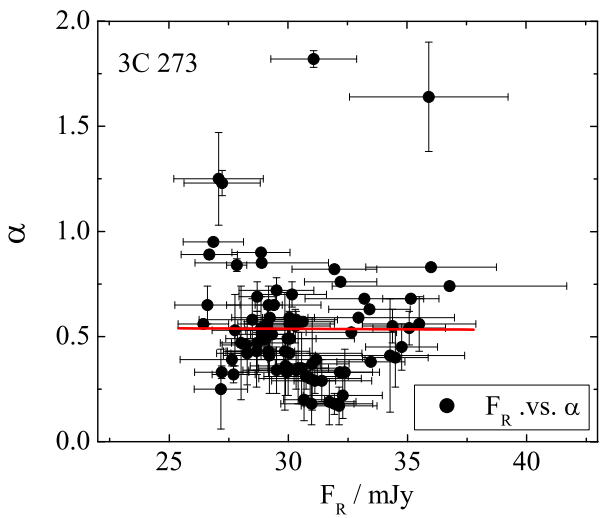
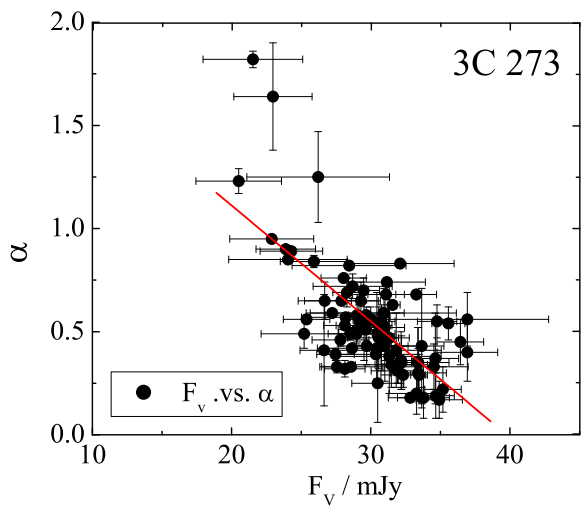
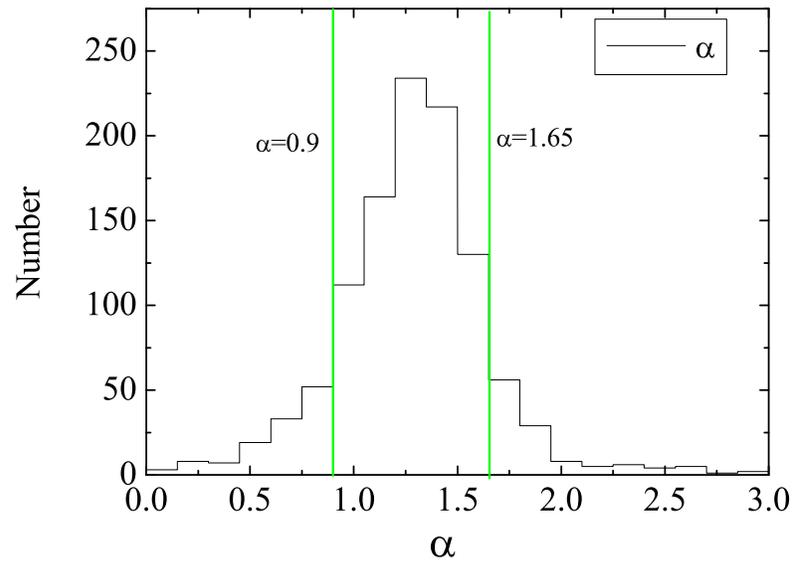
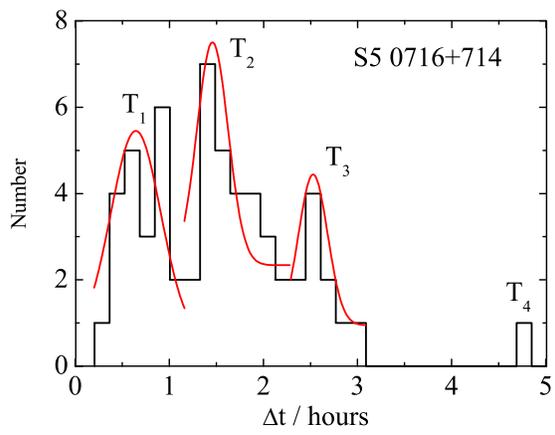
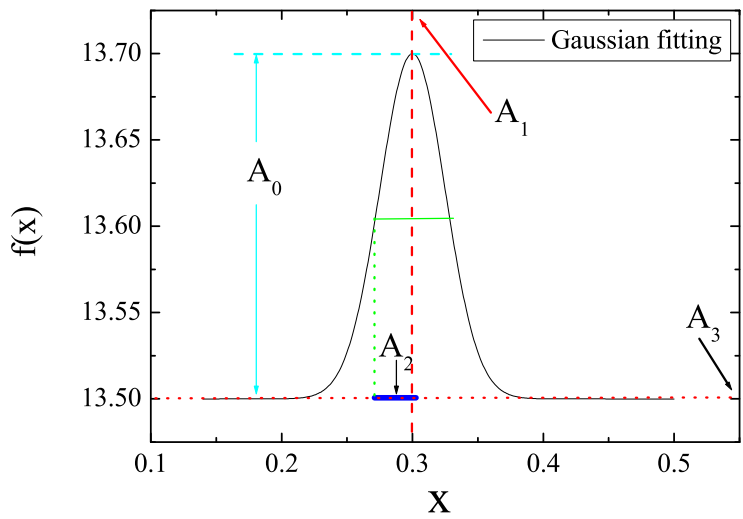
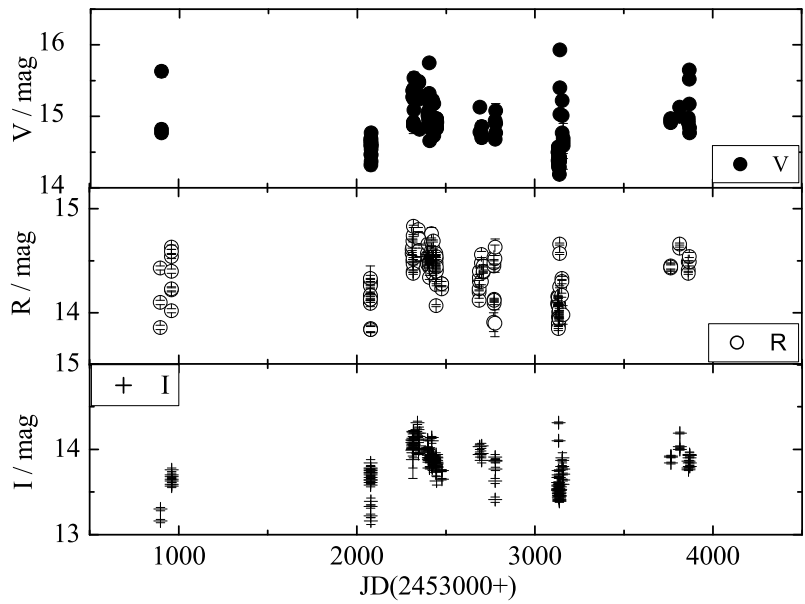
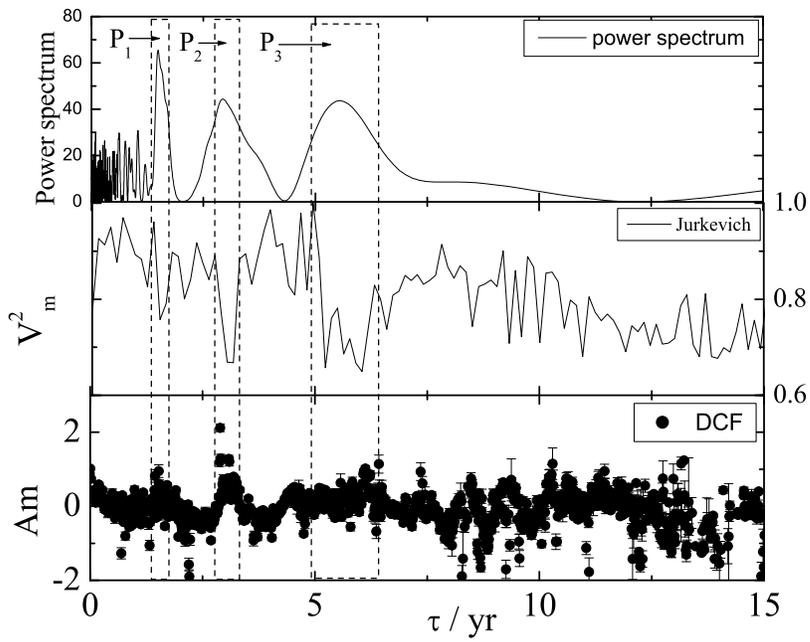
\end{document}